\documentclass{rmf-d}
\usepackage{nopageno,rmfbib,multicol,times,amsmath,amssymb,cite}
\usepackage[T1]{fontenc} 
\usepackage[]{caption2}
\usepackage{graphicx}
\usepackage{blindtext}
\usepackage{color}
\usepackage{hyperref}
\def\rmfcornisa{ Accepted for publication in Revista Mexicana de F\'{\i}sica \hfill}
\def\rmfcintilla{{ Accepted in \it Rev.\ Mex.\ Fis.\/} {\bf }  }
\clearpage \rmfcaptionstyle 
\begin{document}
\markboth{    }{  }

%
%
\title{ QCD Phase Diagram for Large $N_f$ : Analysis from Contact Interaction Effective Potential}
\vspace{-6pt}
\author{Aftab Ahmad }
\address{Institute of Physics, Gomal University, 29220, D.I. Khan, Khyber Pakhtunkhaw, Pakistan. }
\address{Email: aftabahmad@gu.edu.pk}

%
%
\author{ }
\address{ }
\author{ }
\address{ }
\author{ }
\address{ }
\author{ }
\address{ }
\author{ }
\address{ }
\maketitle
%
%
\begin{abstract}
\vspace{1em} 
%
%
In this paper, we discuss the impact of a higher number of light quark flavors, $N_f$, on the QCD phase diagram under extreme conditions. Our formalism is based on the Schwinger-Dyson equation, employing a specific symmetry-preserving vector-vector flavor-dressed contact interaction model of quarks in Landau gauge, utilizing the rainbow-Ladder truncation. We derive expressions for the dressed quark mass $M_f$ and effective potential $\Omega^{f}$ at zero, at finite temperature $T$ and the quark chemical potential $\mu$. The transition between chiral symmetry breaking and restoration is triggered by the effective potential of the contact interaction, whereas the confinement and deconfinement transition is approximated from the confinement length scale $\tilde{\tau}_{ir}$.
Our analysis reveals that at $(T = \mu = 0)$, increasing $N_f$ leads to the restoration of chiral symmetry and the deconfinement of quarks when $N_f$ reaches its critical value, $N^{c}_{f} \approx 8$. At this critical value, In the chiral limit ($m_f = 0$), the global minimum of the effective potential occurs at the point where the dressed quark mass approaches zero ($M_f \rightarrow 0$). However, when a bare quark mass of $m_f = 7$ MeV is introduced, the global minimum shifts slightly to a nonzero value, approaching $M_f \rightarrow m_f$. At finite $T$ and $\mu$,  we illustrate the QCD phase diagram in the $(T^{\chi,C}_{c} -\mu)$ plane, for various numbers of light quark flavors, noting that both the critical temperature $T_c$ and the critical chemical potential $\mu_c $ for chiral symmetry restoration and deconfinement decrease as $ N_f $ increases. Moreover, the critical endpoint  $(T_{EP}, \mu_{EP})$  also shifts to lower values with increasing $N_f $.  Our findings are consistent with other low-energy QCD approaches.
\vspace{1em}
\end{abstract}
\keys{ \bf{\textit{Schwinger-Dysons equation, chiral symmetry breaking, confinement, finite temperature and density,  QCD phase diagram
}} \vspace{-8pt}}
\begin{multicols}{2}

\section{Introduction}
In the past few years, the study of low-energy quantum chromodynamics (QCD) for large number light quark flavors $N_f$ yields important consequences on the chiral symmetry breaking/ restoration and confinement/deconfinement phase transition in the presence of heat bath and external background fields. It has been clearly demonstrated in the Lattice QCD simulation~\cite{LSD:2014nmn,Hayakawa:2010yn,Cheng:2013eu,Hasenfratz:2016dou,LatticeStrongDynamics:2018hun}, the Schwinger-Dyson equation approaches~\cite{bashir2013qcd,Ahmad:2020jzn,Ahmad:2020ifp,Ahmad:2022hbu,Ahmad:2023mqg}, some continuum approaches \cite{Appelquist:1998rb,Hopfer:2014zna,Doff:2016jzk,Binosi:2016xxu,Evans:2020ztq, Zierler:2023qvz} and the NJL model calculation~\cite{Ahmad:2022hbu} that as we increase the number of light quark flavors, the chiral symmetry restores and quarks become deconfined at and above some critical value $N_{f}^{c}$. The critical value must be less than the upper limit of the critical value, denoted as $N_{f}^{c,AF}$, where the asymptotic freedom is believed to exist.
According to~\cite{Politzer:1973fx}, the critical number of flavors for a gauge group $ SU(Nc)$ is given by $( N^{c,AF}_{f} = \frac{11 N_c}{2} )$. For $ N_c = 3$, this yields a critical value of $16.5$. Consequently, QCD is considered conformal in the infrared regime, attributed to the presence of an infrared fixed point which refers to a specific condition in which the beta functions of the QCD couplings are equal to zero~\cite{Caswell:1974gg,Banks:1981nn,gies2006chiral,Appelquist:2007hu,hasenfratz2010conformal,Aoki:2011rrd,Evans:2020ztq} . The range of fermion flavors $(N^{c}_{f} \leq N_f < N^{c,AF}_{f})$ is commonly referred to as the "conformal region"~\cite{Appelquist:2009ty, LSD:2009yru,Zhitnitsky:2013wfa,Lee:2020ihn}. As we approach the upper limit $(N_f< N^{c,AF}_{f})$ of this region, the infrared fixed point is located in a weakly interacting regime. In contrast, at the lower region $(N_f\leq N^{c}_{f})$, the infrared fixed point shifts towards the strongly interacting region, where the coupling becomes progressively stronger as $ N_f $ decreases. As a result, the system enters a phase marked by the breaking of chiral symmetry and the confinement of quarks. \\Furthermore, the study of a large number of light quark flavors $N_f$ plays a significant role in light hadron physics, see for example, recent Lattice QCD studies~\cite{LatticeStrongDynamics:2023bqp} and Schwinger-Dyson studies in  Ref.~\cite{Ahmad:2024emu} discusses the properties of pions and kaons across various light quark flavors in detail. This research reveals that at and above the critical $N_f \approx 8$, the masses of the pseudoscalar mesons increase, indicating that chiral symmetry is restored and mesons behave like free particles.
Additionally, the Schwinger quark-antiquark pair production rate is sensitive to the increasing number of light quark flavors. It has been demonstrated that, in the presence of a pure electric field, the pair production rate accelerates with a higher number of flavors~\cite{Ahmad:2023mqg}. In Ref.~\cite{Ahmad:2020ifp}, a study of large $N_f$ in a heat bath and magnetic field background shows that there is a critical number of flavors, $N^{c}_f \approx 8$, at which chiral symmetry is restored and deconfinement occurs, while the critical temperature $T_c$ decreases as $N_f$ increases.
Extending the study of large $N_f$ to finite temperature $T$ and chemical potential $\mu$ provides deeper insights into the nature of the chiral phase transition within the QCD phase diagram. A model-based study, such as the Nambu-Jona-Lasinio (NJL) model, predicts that increasing the number of light quark flavors, $N_f$, leads to a decrease in all phase diagram parameters, including the critical temperature $T_c$, the critical quark chemical potential $\mu_c$, and the critical endpoint $(T_E, \mu_E)$, as discussed in Ref.~\cite{Ahmad:2022hbu}.
In this work, we focus on studying the QCD phase diagram in the $( T^{\chi,C}_{c}-\mu )$ plane for various numbers of light quark flavors  $N_f$. 
 This exploration is crucial for understanding the phase transitions that hadronic matter underwent in the early stages following the Big Bang. Specifically, we examine the transition from hadronic matter to quark-gluon plasma~\cite{rischke1988phase}, quarkyonic matter \cite{mclerran2009quarkyonic,McLerran:2018hbz,Bluhm:2024uhj}, neutron star formation~\cite{shao2011evolution, Adhikari:2024bfa}, and the color-flavor locked (CFL) region in the QCD phase diagram \cite{Barrois:1977xd,Casalbuoni:1999zi,Rajagopal:1999cp}. Recent advancements in detectors at various research centers, including the sPHENIX detector and the complementary STAR upgrades at RHIC, as well as enhancements at ALICE, ATLAS, CMS, and LHCb, have ushered in a new multi-messenger era for hot quantum chromodynamics (QCD)~\cite{Arslandok:2023utm}). This era leverages the combined constraining abilities of low-energy hadrons, jets, thermal electromagnetic radiation, heavy quarks, and exotic bound states. Additionally, the increased luminosity at the LHC, alongside other experimental facilities like RHIC and the Compact Baryonic Matter (CBM) experiments, and new facilities under construction in FAIR~\cite{durante2019all} and NICA \cite{Kolesnikov:2020qfw}, presents a remarkable opportunity to investigate the phase transition from hadronic matter to quark-gluon plasma and related phenomena.\\ 
 It is well known that at temperatures near zero, color-singlet hadrons are generally considered the basic building blocks of low-energy  QCD. However, when the temperature surpasses a critical threshold $T_c $, the interactions weaken, causing hadrons to transition into a new phase. In this phase, quarks and gluons emerge as the new fundamental components, chiral symmetry is restored, and quarks become deconfined. Lattice QCD calculations~\cite{Aoki:2006we, Cheng:2006qk, Bhattacharya:2014ara, deForcrand:2014tha, HotQCD:2018pds, Borsanyi:2020fev, Guenther:2020jwe, Borsanyi:2025lim}, Schwinger-Dyson equations~\cite{Qin:2010nq, Fischer:2011mz, gutierrez2014qcd, Eichmann:2015kfa, Ahmad:2015cgh, Gao:2016qkh,Ahmad:2016iez, Fischer:2018sdj, Shi:2020uyb, Ahmad:2020ifp, Ahmad:2020jzn}, and various effective models of non-perturbative QCD~\cite{klevansky1992nambu, buballa2005njl, costa2010phase, Ayala:2011vs,Marquez:2015bca , Ahmad:2015cgh, Ayala:2017gek, Ayala:2021nhx, Ahmad:2022hbu,Ahmad:2023ecw} all indicate that the transition in question manifests as a crossover when a finite current quark mass $m$ is considered. Conversely, calculations in the chiral limit reveal a second-order phase transition. Nonetheless, as the quark chemical potential $\mu$ increases, this same physical behavior continues to hold. The nature of the phase transition shifts from a crossover to a first-order transition at the critical endpoint (CEP) in the QCD phase diagram, typically depicted on the $( T_c-\mu )$ plane. The precise location of this critical endpoint remains elusive, sparking significant scientific interest and prompting experimental designs aimed at its observation. However, various non-perturbative QCD model and Lattice QCD simulation  suggest it lies within the range $(\mu_E/T_c = 1.0 - 2.0, T_E/T_c = 0.4 - 0.9)$ for two-flavor QCD~\cite{Sasaki:2007qh,Costa:2008yh,Fu:2007xc,Abuki:2008nm,Loewe:2013zaa,Kovacs:2007sy,Schaefer:2007pw,gutierrez2014qcd,
Bazavov:2011nk,Ahmad:2015cgh,Ahmad:2022hbu,fodor2002lattice,gavai2005critical,li2009study,deForcrand:2006ec}.\\
In this study, we aim to investigate the phenomena of chiral symmetry breaking and restoration, as well as the confinement and deconfinement phase transitions, particularly focusing on a large number of light quark flavors, $N_f$. Our objective is to map out the QCD phase diagram. To accomplish this, we employ a confining variant of the Nambu-Jona-Lasinio (NJL) model. This model features a symmetry-preserving vector-vector contact interaction among quarks and incorporates multiple light quark flavors $N_f$ \cite{Ahmad:2020jzn}. We analyze this within the framework of Schwinger-Dyson equations, utilizing the Landau gauge and a Schwinger optimal time regularization method, while considering finite temperature $T$ and chemical potential $\mu$. 
In our analysis, the gap equation derived from this model can be integrated concerning the dynamically generated mass, thereby defining the effective thermodynamic potential of the current contact interaction model~\cite{Ahmad:2023mqg}. 
The critical  number of flavors $N^{c}_{f}$, the critical temperature $T^{\chi,C}_{c}$ and the critical chemical potential $\mu^{\chi,C}_{c}$ of chiral symmetry breaking and restoration, as well as the confinement-deconfinement phase transition, can be approximated from the  effective contact interaction thermodynamical potential $\Omega$ and the confinement length scale $\tilde{\tau}_{ir}$ (see for detail discussion~\cite{Ahmad:2023ecw}, and also
~\cite{Wang:2013wk, Ahmad:2016iez, Ahmad:2020ifp,Ahmad:2020jzn}), respectively. 
 Notably, in this model, chiral symmetry restoration and deconfinement occur simultaneously~\cite{Marquez:2015bca, Ahmad:2016iez, Ahmad:2020ifp, Ahmad:2020jzn}.\\
The remainder of this manuscript is organized as follows: In Section 2, we introduce the general formalism for the gap equation using contact interaction model and the contact interaction effective potential in vacuum for large $N_f$. Section 3 delves into the gap equation and effective potential at finite temperature and chemical potential for large$N_f$. In Section 4, we present numerical solutions for the gap equation and effective potential, and we sketch the phase diagram in the $(T^{\chi,C}_{c}-\mu)$ plane for various $N_f$. In Section 5, we summaries our findings.
\section{$N_f$-flavors dressed contact interaction model gap equation and  the  effective potential in vacuum}\label{section-II} 
The Schwinger-Dyson equation for the  flavor-dressed quark  propagator  in the vacuum is given by:
\begin{eqnarray}
S_{f}^{-1}(p)&=S^{-1}_{0,f}(p) + \Sigma_{f}(p)\,.\label{CI1}
\end{eqnarray}
Here, $S_{0,f}(p)=(\gamma\cdot p- m_{f} + i\epsilon)^{-1}$ stands for the bare quark propagator in Minkowski space, the subscript $f$ is denotes the quark flavors  and   $\Sigma(p)$ is the self energy given by: 
\begin{eqnarray}
\Sigma_{f}(p)=-i\int \frac{d^4k}{(2\pi)^4} g^{2}
 \Delta_{\mu\nu}(q)\frac{\lambda^a}{2}\gamma_\mu S(k)
\frac{\lambda^a}{2}\Gamma_\nu(p,k)\,.\label{CI2}
\end{eqnarray}
Where, $\Gamma_\nu (k,p)$ is the dressed quark-gluon vertex, $g^{2}$ is QCD coupling constant and  $$\Delta_{\mu\nu}(q) = -i\frac{\mathcal{G}(q)}{q^2}\left(g_{\mu\nu} -\frac{q_{\mu} q_{\nu}}{q^2}\right),$$  is the gluon propagator (in the Landau gauge). Here $g_{\mu \nu}$ is the metric tensor (in Minkowski space), $q=k-p$ is the gluon four-momentum and  $\mathcal{G}(q)$ is the gluon dressing function.  The symbol $m_f$  stands for bare light quark mass and in the chiral limit $m_f=0$. Here $\lambda^a$'s   are the Gell-Mann matrices, and  in the $SU(N_c)$ representation these matrices satisfies the following identity:
\[
\sum^{N^{2}_{c}-1}_{a=1}\frac{{\lambda}^a}{2}\frac{{\lambda}^a}{2}=\frac{1}{2}\left(N_c - \frac{1}{N_c} \right)I, \label{CI3}
\] 
here, $N_c$ represents the number of colors and $I$  is the identity matrix. In the present scenario, we use the rainbow-Ladder truncation i.e., $\Gamma_\nu (k,p)=\gamma_{\nu}$. 
The $N_f$-flavors dressed form of the symmetry-preserving contact interaction model~\cite{Ahmad:2020jzn, Ahmad:2022hbu,Ahmad:2024emu} is given by:
\begin{eqnarray}
  g^2 \frac{\mathcal{G}(q)}{q^2}\Bigg|_{q\rightarrow 0}= \frac{4 \pi
  \alpha_{\rm Ir}} {\mathcal{M}_{g}^{2}} \sqrt{1 - \frac{(N_{f}-2)}{\mathcal{Z}_{f}^{c}}}= \alpha_{\rm eff}(N_f) \,.\label{CI4}
\end{eqnarray}
Where $\alpha_{\rm ir}=0.93\pi$, is the strength parameter for the infrared-enhanced interaction and   $\mathcal{M}_g=800$ MeV is the dynamically generated gluon mass scale in the infrared region~\cite{Boucaud:2011ug}. The $\mathcal{Z}_{f}^{c}=9.98$, represents  the guessed critical number of flavors, as discussed in detail in~\cite{Ahmad:2020jzn,Ahmad:2022hbu}.
In this  particular model truncation, the dynamical quark mass function remains momentum independent, and the dressed quark propagator can be expressed as ~\cite{Solis:2019fzm,Ahmad:2022hbu}~:
\begin{eqnarray}
S_{f}(k)=\frac{\gamma\cdot k+ M_{f} }{k^2- M_{f}^{2} + i\epsilon}\label{CI5}.
\end{eqnarray}
Here, $M_f$ is the dressed or effective quark mass.
Inserting Eqns.~(\ref{CI2}) -(\ref{CI5}) into Eq.(\ref{CI1}), using $N_c=3$  and
simplifying, we have
\begin{eqnarray}
M_f = m_f +\frac{16i\alpha_{\rm eff}(N_f)}{3}  \int\frac{d^{4}k}{(2\pi)^4}\frac{M_f}{k^2 - M_{f}^{2} + i\epsilon}.
\label{CI6}
\end{eqnarray}
Now, we can split  the four-momentum and four-dimensional momentum integral  into time and
space components. The space  part of the four momentum  is denoted by  $\textbf{k}$ and the temporal  part by $k_0$. So, Eq.~(\ref{CI6}) can be written as
\begin{eqnarray}
M_f  = m_f  + \frac{16i\alpha_{\rm eff}(N_f)}{3} \int_{0}^{\infty}\frac{d^3\textbf{k}}{(2\pi)^4} \int_{-\infty}^{+\infty} \frac{M_f dk_{0}}{ {k_{0}}^{2} - E_{k}^{2} + i\epsilon}.
\label{CI8}
\end{eqnarray}
 After  performing integration  over the time component of Eq.~(\ref{CI8}), we  have:
\begin{eqnarray}
M_f  = m_f  +\frac{16i\alpha_{\rm eff}(N_f)}{3}  \int_{0} ^{\infty} \frac{d^3 \textbf{k}}{(2\pi)^4} \frac{\pi M_f}{i E_{k}}.
\label{CI10}
\end{eqnarray}
Here, $E_{k}=\sqrt{{\textbf{k}}^2 + M_{f}^{2}}$ represents the energy per particle.
Using $d^3 \textbf{k} = \textbf{k}^2 d\textbf{k} \sin\theta d\theta d\phi$ and upon taking the angular integration,  Eq.~(\ref{CI10}) can be written as:
\begin{eqnarray}
M_f  =  m_f  + \frac{4\alpha_{\rm eff}(N_f)}{3\pi^2} \int_{0} ^{\infty}d \textbf{k} \frac{\textbf{k}^2}{\sqrt{\textbf{k}^2 +M_{f }^{2}}}.
\label{CI11}
\end{eqnarray}
The integral in Eq.~(\ref{CI11}) is diverging integral and needs to  be regularized. Here, we  use  the Schwinger proper time regularization procedure, by using the following identity:
\begin{eqnarray} 
\frac{1}{a^n}=\frac{1}{\Gamma(n)}\int^{\infty}_{0} d\tau \tau^{n-1}e^{-\tau a},\label{CI12} 
\end{eqnarray} 
where $\Gamma (n)$ is the Gamma function. We use $a=\textbf{k}^2 + M_{f}^{2}$ and $n=1/2$ and introducing the cut-offs $\tau_{ir}=1/\Lambda_{ir}$ along with an ultraviolet cut-off~$\tau_{uv}=1/\Lambda_{uv}$~\cite{Ebert:1996vx,GutierrezGuerrero:2010md,Roberts:2011cf,Ahmad:2023ecw}, we have:
\begin{eqnarray} 
\frac{1}{\sqrt{\textbf{k}^2 + M_{f}^{2}}}&=&\int^{\infty }_{0} \frac{d\tau e^{-\tau(\textbf{k}^2+M_{f}^{2})}}{\sqrt{\pi \tau}}\rightarrow\int_{\tau^{2}_{uv}}^{\tau^{2}_{ir}} \frac{d\tau e^{-\tau(\textbf{k}^2+M_{f}^{2})}}{\sqrt{\pi \tau}} 
\nonumber\\
&=&\frac{{\rm erf}(x_{ir}) - {\rm erf}(x_{uv})}{\sqrt{k^2 + M_{f}^{2}}}.
\label{CI13}
\end{eqnarray} 
Where ${\rm erf}(x_{ir, uv}=\sqrt{k^2 + M_{f}^{2}} \, \tau_{ir,uv})$  are the error functions, i.e., $${\rm erf}(x)=\frac{2}{\sqrt{\pi}}\int^{x}_{0}\frac{e^{-t^2}}{t}dt.$$ 
The pole in Eq.~(\ref{CI13}), located at $\textbf{k}^2=-M_{f}^{2}$ disappears from the quark propagator when both the numerator and denominator vanish at that point. The ultraviolet regulator, $\tau_{uv}=\Lambda^{-1}_{uv}$, sets the scale for dimensional quantities dynamically. The infrared regulator, $\tau_{ir}=\Lambda^{-1}_{ir}$ where , with a non-zero value, aids in interpreting confinement~\cite{Ebert:1996vx, Roberts:2011cf, Roberts:2011wy} and is often known as the confinement length scale ~\cite{Wang:2013wk, Ahmad:2016iez, Ahmad:2020ifp, Ahmad:2020jzn,Ahmad:2023ecw,Ahmad:2023mqg}. Analysis of Eq.~(\ref{CI13}), reveals that the propagator lacks real or complex poles, aligning with the concept of confinement.  Essentially, an excitation described by a pole-less propagator cannot reach its mass-shell~\cite{Ebert:1996vx}. By substituting Eq.~(\ref{CI13}) in 
Eq.~(\ref{CI11}) and integrating over $\textbf{k}$, the gap equation Eq.~(\ref{CI11}) for the dressed mass is simplified to:
\begin{eqnarray}
 M_f =  m_f +\frac{4\alpha_{\rm eff}(N_f) M_f }{3\pi^2} \bigg( \frac{e^{-M_f^2 \tau_{ir}^{2}}}{\tau_{ir}^{2}} + \frac{e^{-M_f^2 \tau_{uv}^{2}}}{\tau_{uv}^{2}}-\nonumber\\
M_f^2 {\rm Ei}(-M_f^2 \tau_{ir}^{2}) + M_f^2 {\rm Ei}(-M_f ^2 \tau_{uv}^{2})\bigg),\label{CI14}
\end{eqnarray}
with ${\rm Ei}(x)=\int^{x}_{-\infty}\frac{e^{-t}}{t}dt,$ is the exponential integral function. 
The quark-antiquark condensate in the present case is defined as:
\begin{eqnarray}
-\langle \bar{q} q\rangle= \frac{M_f -m_f }{\alpha_{\rm eff}(N_f)}.\label{CI15}
\end{eqnarray}
The $N_f$-flavor dressed effective contact interaction potential in vacuum can be obtained by re-arranging  Eq.~(\ref{CI14}) and upon integration over $M_f$~\cite{Ahmad:2023ecw,Ahmad:2023mqg}, is thus, given by:
\begin{eqnarray}
\Omega^{f}(M,N_f)
= \Omega_{0}+ \Omega_{vac},
\label{CI15a}
\end{eqnarray}
here, the $\Omega_0$ is related to the  square of condensate  and is given by
\begin{eqnarray}
    \Omega_{0}=\frac{(M_f-m_f)^2}{2\alpha^{N_c}_{\rm eff}(N_f)},\label{CI15b}
\end{eqnarray}
and 
\begin{eqnarray}
\Omega_{vac}=\frac{1}{12\pi^{2}} \bigg( \frac{e^{-M_{f}^{2} \tilde{\tau}_{ir}^{2}} (1 - M_{f}^{2} \tilde{\tau}_{ir}^{2})}{\tilde{\tau}_{ir}^{4}}+\nonumber\\ \frac{e^{-M_{f}^{2} \tau_{uv}^{2}} (-1 + M_{f}^{2} \tau_{uv}^2)}{\tau_{uv}^{4}}  - M_{f}^{4} {\rm Ei}(-M_{f}^{2} \tilde{\tau}_{ir}^2) + \nonumber\\M_{f}^{4}{\rm Ei}(-M_{f}^{2} \tau_{uv}^{2}) \bigg)+const.,\label{CI16}
\end{eqnarray}
is the regularized vacuum part of the flavor-dressed  effective potential. This indicates  that the state with the lowest value of $\Omega^{f}$, achieved by fulfilling the conditions $ \partial \Omega^{f}/\partial M_f = 0 $ and $\partial^2 \Omega^f/\partial M_{f}^{2} \geq 0$, is regarded as the most stable. In this context, we define the confinement scale as
$$\tilde{\tau}_{ir}(M, N_f) = \tau_{ir}  \frac{M_f(2,0,0)}{M_f(N_f,0,0)},$$
where $M_f(N_f,0,0)$ represents the dressed mass for a general number of flavors $N_f$, and $ M_f(2,0,0)$ corresponds to $N_f = 2$, both evaluated at zero temperature $T = 0 $ and zero chemical potential $ \mu = 0$. In the following section, we will examine the gap equation and the effective potential under finite temperature and chemical potential conditions.
\section{$N_f$-flavors dressed gap equation and  contact interaction effective potential at finite $T$ and $\mu$}
 The gap equation Eq.~(\ref{CI6}), at finite temperature $T$ and quark chemical potential $\mu$ ,  can be obtained  by using the following  convention for momentum integration:
\begin{eqnarray}
\int\frac{d^4k}{i(2\pi)^4} f(k_0,{\bf{k}})\rightarrow T \sum_{n} \int\frac{d^3 k}{(2\pi)^3}f(\tilde{\omega}_n,\bf{k}), \label{CI17}
\end{eqnarray}
here $\tilde{\omega}_n = i(2n+1)\pi T+\mu$, stands for the fermionic Matsubara frequencies. Inserting Eq.~(\ref{CI17}) in Eq.~(\ref{CI6}) and simplifying, we have
\begin{eqnarray}
M_f = m_f + \frac{4\alpha_{\rm eff}M_f }{3\pi^2}\int^{\infty}_{0} \frac{d^{3}\textbf{k}}{(2\pi)^{3}}\frac{1}{\sqrt{\textbf{k}^2 + M_{f}^{2}}}\nonumber\\(1- n_{F}(T, \mu)+\bar{n}_{F} (T, \mu))).
\label{CI18}
\end{eqnarray}
Where  $n_F(T, \mu)$ and $\bar{n}_F(T, \mu)$ represents the Fermi occupation numbers for the quarks and antiquarks, respectively, can be defined as: 
\begin{eqnarray}
n_{F}(T, \mu) = \frac{1}{e^{y_{-}}+1} ,\qquad \bar{n}_{F}(T, \mu)= \frac{1}{e^{y_{+}} + 1},
\label{CI19}
\end{eqnarray}
where $$y_{\mp}=(\sqrt{\textbf{k}^2 + M_{f}^{2}}\mp\mu)/T.$$
By isolating the vacuum component from the medium and applying appropriate time regularization as shown in Eq.~(\ref{CI13}), we can simplify the gap equation in Eq.~(\ref{CI18}) as follows:
\begin{eqnarray}
 M_f= m_f +\frac{\alpha_{ \rm eff}(N_f) M_f}{3\pi^2} \bigg( \frac{e^{-M_{f}^{2} \tilde{\tau}_{ir}^{2}}}{\tau_{ir}^{2}} +  \frac{e^{-M_{f}^{2} \tau_{uv}^{2}}}{\tau_{uv}^{2}}\nonumber\\- M_{f}^{2} {\rm Ei}(-M_{f}^{2} \tilde{\tau}_{ir}^{2}) + M_{f}^{2} {\rm Ei}(-M_{f}^{2} \tau_{uv}^{2})\bigg) 
\nonumber\\
-\frac{4\alpha_{\rm eff} M_{f}}{3\pi^2}\int_{0}^{\infty} d\textbf{k} \frac{\textbf{k}^2}{\sqrt{\textbf{k}^2 + M_{f}^{2}}} \left[ n_{F}(T, \mu)+ \bar{n}_{F} (T, \mu)\right],
\label{CI20}
\end{eqnarray}
In this context, we define $$\tilde{\tau}_{ir} = \tau_{ir} \frac{M(2,0,0)}{M(N_f,T,\mu)},$$ where $M_{f}(N_f,T,\mu)$ denotes the dressed mass influenced by the medium. This formulation indicates that as chiral symmetry is restored, $\tilde{\tau}_{ir}$ approaches infinity, suggesting a simultaneous onset of deconfinement.
Crucially, it’s important to highlight that since the plasma does not affect the ultraviolet dynamics, the medium component of the gap equation remains unaffected by regularization. Only the vacuum component necessitates such regularization. Integrating Eq.~(\ref{CI20}) with recept to dressed mass $M_f$ and arranging the terms  the expression for flavor-temperature-chemical potential effective thermodynamical potential $\Omega^{f}(T,\mu)$ is given by: 
\begin{eqnarray}
\Omega^{f}(T,\mu)&=&\Omega_{0}+\Omega_{vac}+\Omega_{med},
\label{CI21}
\end{eqnarray}
here, the medium part of the effective potential is: 
\begin{eqnarray}
\Omega_{med}=
-\frac{4}{3\pi^{2}}\int_{0}^{\infty}d\textbf{k} \textbf{k}^2 \bigg(T \log\left[1 + e^{-y_{-}}\right] \nonumber\\
+ T \log\left[1 + e^{-y_{+}}\right]\bigg).\label{CI22}
\end{eqnarray}
The equation  Eq.~(\ref{CI20}), presented above aligns with the gap equation derived in Eq.~(\ref{CI15a}) when we set $T=\mu=0$.  

 \section{Numerical Results}\label{section-III}
\begin{figure}[H]
\begin{center}
\includegraphics[width=0.4\textwidth]{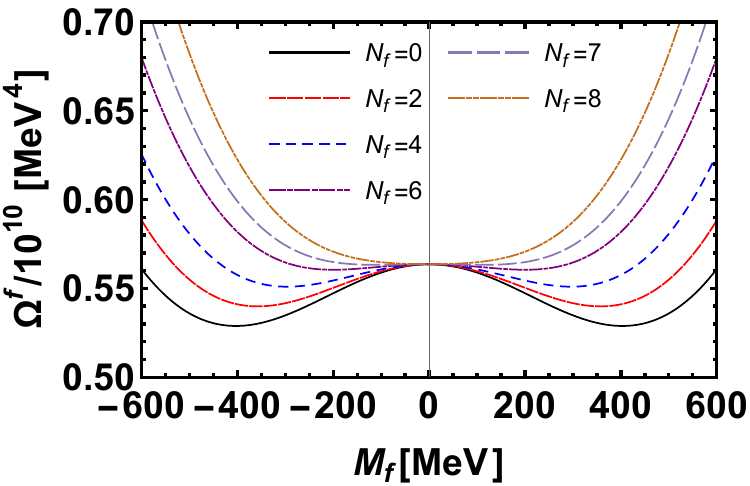}
\includegraphics[width=0.4\textwidth]{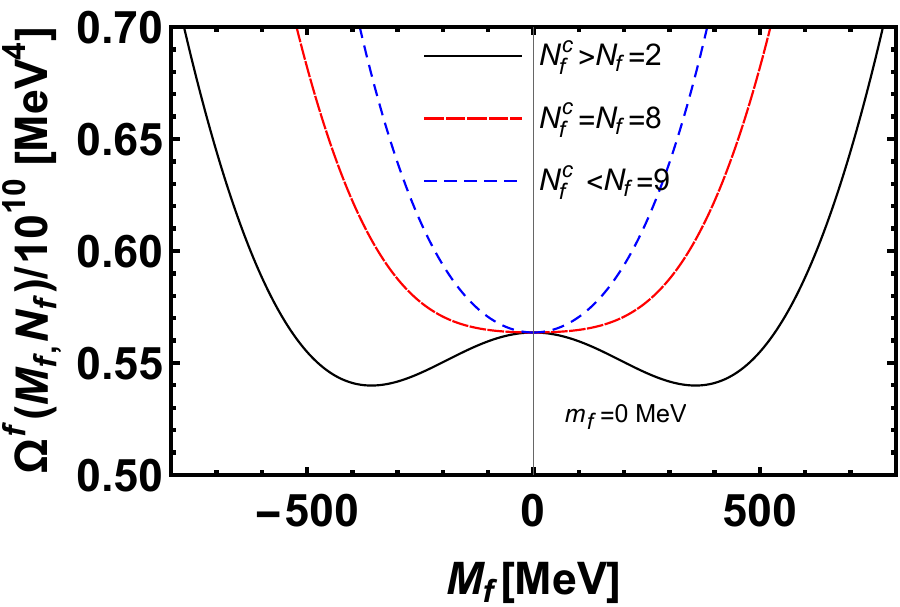}
\caption{ 
\textit{Upper panel}: The behavior of the contact interaction effective potential for various numbers of light quark flavors $N_f$ shows that  the effective potential has the stable  global  at $M\neq0$ for lower values of  $N_f$  indicate  chiral symmetry breaking in the chiral limit. As $N_f$ increases, particularly for $N_f \geq N^{c}_{f} \approx 8$, the minima shift towards $M_f = 0$, indicating chiral symmetry restoration phase.  \textit{Lower panel}: The behavior of the effective potential below, at and above the critical number of flavors $N^{c}_{f}$. This plot shows that  for $N_{f}=2< N^{c}_{f}$, the global minimum is positioned at non-zero  values of $M_f$ ( $M_f\pm358$), representing the chiral symmetry broken phase. For $N_{f}= N^{c}_{f}$, the chiral symmetry restored and the minima shifted towards  $M_f\rightarrow 0$. For  $N_{f}=9>N^{c}_{f}$, the effective potential has stable solution at   $M_f=0$, representing the the chiral symmetry restoration phase.There is no unstable solution (or more then one global minimum) in the effective potential from chiral symmetry broken to restoration phase, so the nature of the transition is continuous and second-order in the chiral limit.}
\label{Fig1}
\end{center}
\end{figure}

\begin{figure}[H]
\begin{center}
\includegraphics[width=0.4\textwidth]{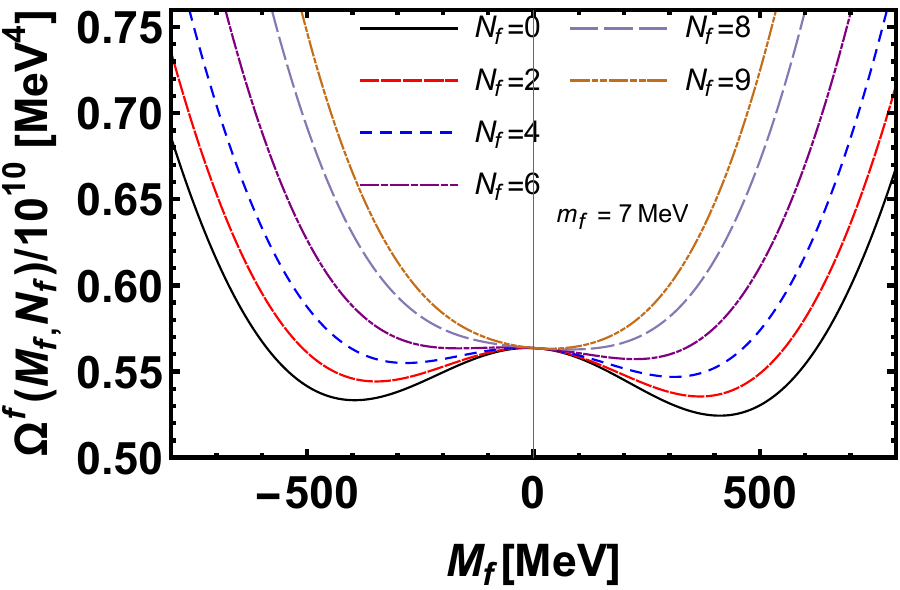}
\includegraphics[width=0.4\textwidth]{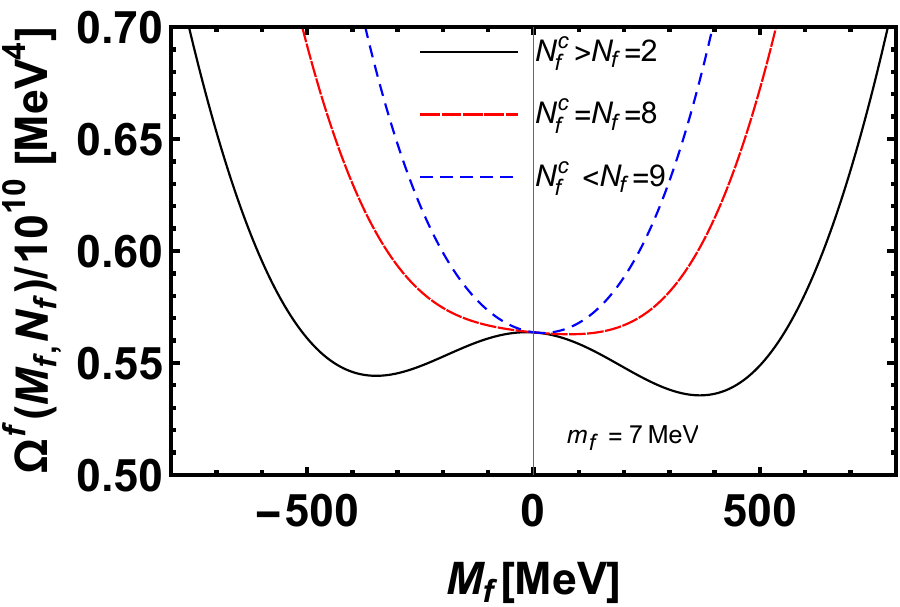}
\caption{  \textit{Upper panel}:
 The contact interaction effective potential for various numbers of light quark flavors $N_f$ with current quark mass $m_f=7$ MeV. The stable global minima in the effective potential for lower values of $N_f$ are located at $M_f>m_f$  indicate that the   chiral symmetry is broken. As $N_f$ increases, particularly for $N_f \geq N^{c}_{f} \approx 8$, these minima shift towards $M_f\rightarrow m_f$, indicating the partial restoration of chiral symmetry.  \textit{Lower panel}: The behavior of the effective potential below, at and above the critical number of flavors $N^{c}_{f}$ with bare quark mass $m_f$. This plot shows that  for $N_{f}=2< N^{c}_{f}$ the global minimum is positioned at non-zero  values of $M_f$ ( $M_f=\pm368$), representing the chiral symmetry broken phase. For $N_{f}= N^{c}_{f}$, the chiral symmetry restored and the minima shifted towards  $M_f\rightarrow m_f$. For  $N_{f}=9>N^{c}_{f}$, the effective potential has stable solution at   $M_f=m_f$, representing the the chiral symmetry restoration phase. There is no unstable solution in the effective potential from chiral symmetry broken to restoration phase and hence, the nature of the transition is cross-over.}
\label{Fig2}
\end{center}
\end{figure}

\begin{figure}[H]
\begin{center}
\includegraphics[width=0.4\textwidth]{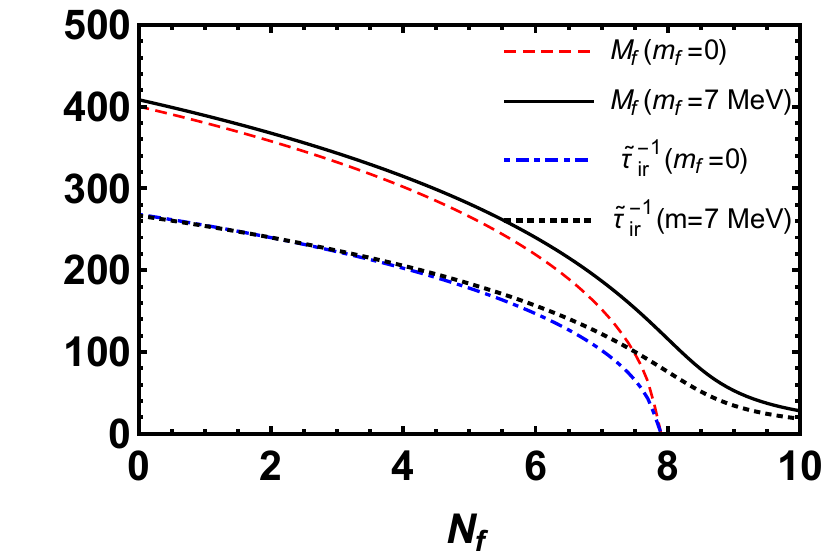}
\caption{ 
Behavior of the dressed mass $M_f$  and confinement length scale $\tilde{\tau}^{-1}_{ir}$ for large number of light quark flavors in the chiral limit and with current quark mass. All parameters monotonically decreasing function of $N_f$. In the chiral limit, at and above  the critical $N^{c}_f\approx8$, the dressed mass and confining length scale vanishes, the   chiral symmetry  is restored and the quark becomes deconfined. Plot with bare quark mass represents that near at or above $N^{c}_f\approx8$, the dress mass vanishes and the only bare quark mass survives, the chiral symmetry is restored through smooth cross-over and quark becomes deconfined.}
\label{Fig3}
\end{center}
\end{figure}

\begin{figure}[H]
\begin{center}
\includegraphics[width=0.4\textwidth]{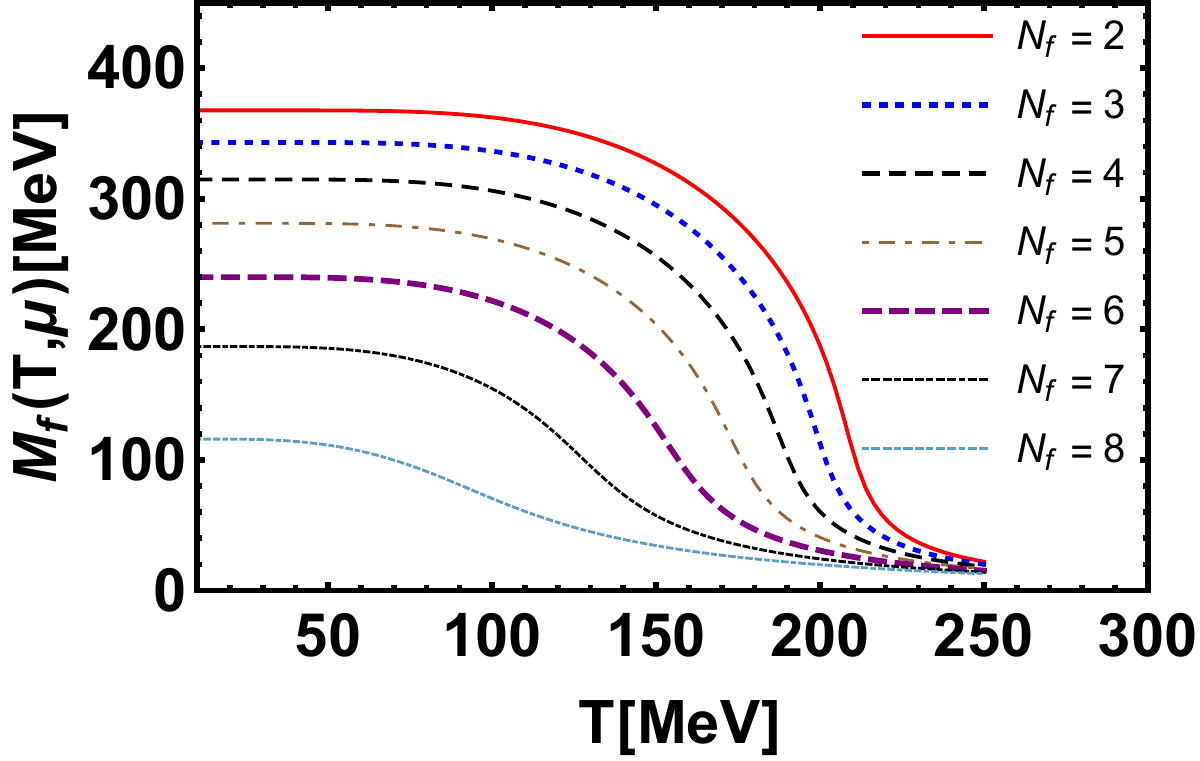}
\caption{ 
Behavior of the  dressed quark  mass  as a function of temperature  for various $N_f$ and  at $\mu=0$. This plot shows the monotonically smooth decreasing behavior of  the dressed mass with a temperature. Upon increasing  $N_f$, it suppresses.}
\label{Fig4}
\end{center}
\end{figure}

\begin{figure}[H]
\begin{center}
\includegraphics[width=0.4\textwidth]{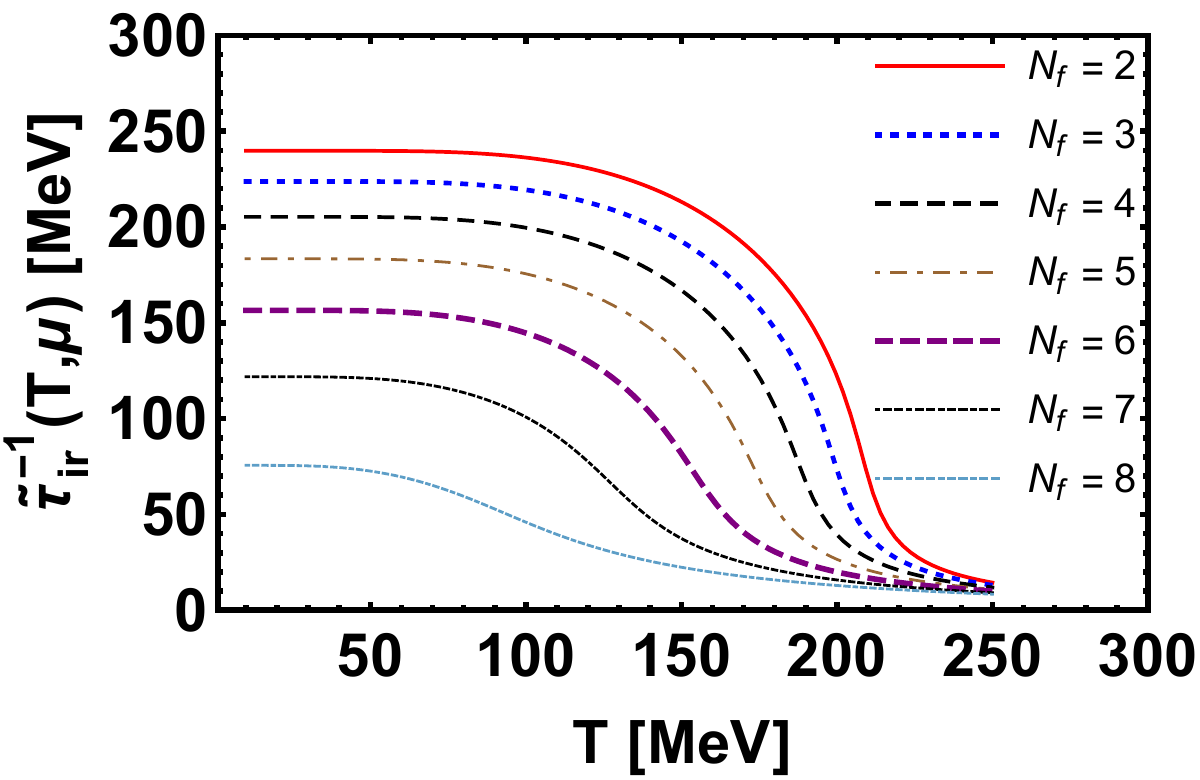}
\caption{ 
The confinement scale $\tilde{\tau}^{-1}_{ir}$ as a function of temperature  for various $N_f$ and $\mu=0$. This plot indicates the monotonically smooth decreasing behavior of  the confinement scale with a temperature. Upon increasing  $N_f$, the confinement scale suppresses.}
\label{Fig5}
\end{center}
\end{figure}

\begin{figure}[H]
\begin{center}
\includegraphics[width=0.4\textwidth]{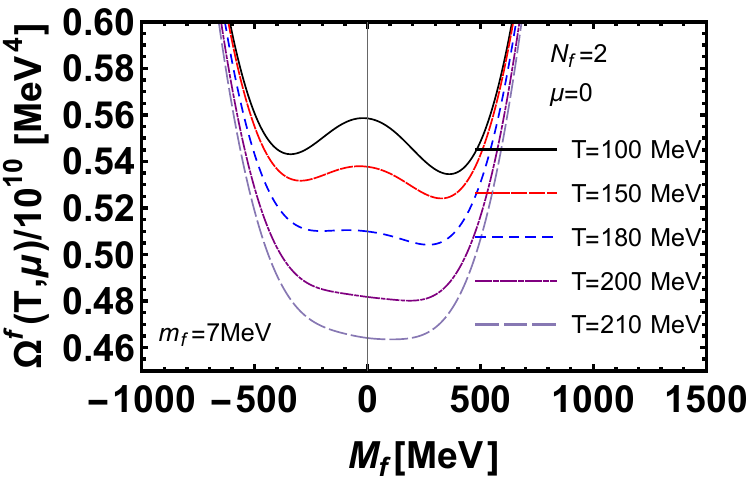}
\includegraphics[width=0.4\textwidth]{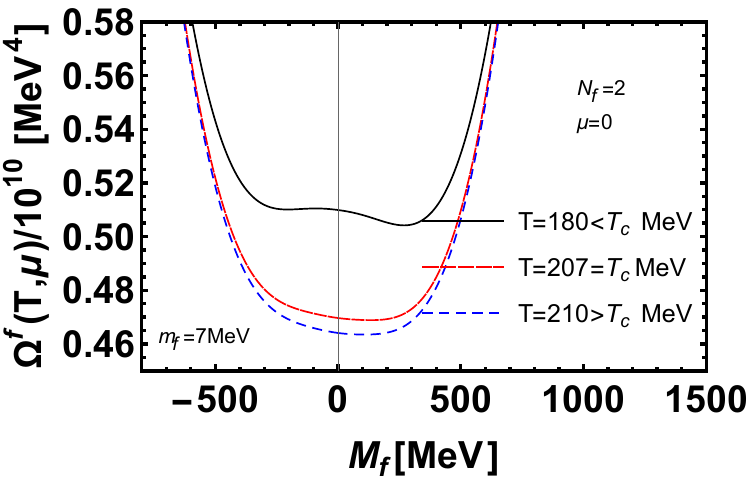}
\caption{ 
\textit{Upper Panel}: the behavior of the effective potential for fixed  $N_f=2$,  for various $T$ and at $\mu=0$. This plot shows that  upon increasing the temperature  the global minima shifted towards the lower values where above $T\approx200$ MeV,  all the minima shift towards $M_f\rightarrow m_f$, where the chiral symmetry is partially restored. \textit{Lower panel}:  Behavior of the effective potential at $(T < T_c)$,~$(T = T_c)$, and $(T > T_c)$. 
It clearly shows that at $(T < T_c)$  the effective potential has a stable global minimum at higher values of $M_f$ below $T_c$, indicating a  chiral symmetry broken phase. At $T = T_c$, this global minimum align with $M = m_f$, where a stable solution exists. For $T > T_c$, the global minimum is positioned at $M_f = m_f$, where chiral symmetry is partially restored.
This plot clearly demonstrates that the effective potential provides stable solutions  from the chiral symmetry-broken phase to the restored phase across all temperature ranges, indicating that the nature of the transition is a crossover.}
\label{Fig6}
\end{center}
\end{figure}

\begin{figure}[H]
\begin{center}
\includegraphics[width=0.4\textwidth]{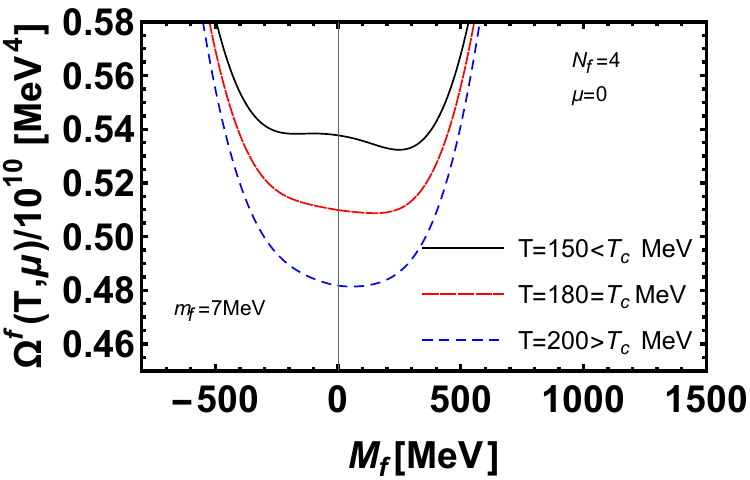}
\caption{ 
The effective potential for fixed  $N_f=4$,  at $(T < T_c)$,~$(T = T_c)$, and $(T > T_c)$. This plot indicates that at  $T=T_c=$ MeV, the global minimum is aligned with $M_f= m_f$ and  at $(T > T_c)$,  it is  located at $M_f= m_f$, which corresponds to the chiral symmetry  restoration phase. However,  for $T<T_c$ , the global minimum is located at $M_f>m_f$, the  chiral symmetry is broken and the transition the its nature is  cross-over.}
\label{Fig7}
\end{center}
\end{figure}

\begin{figure}[H]
\begin{center}
\includegraphics[width=0.4\textwidth]{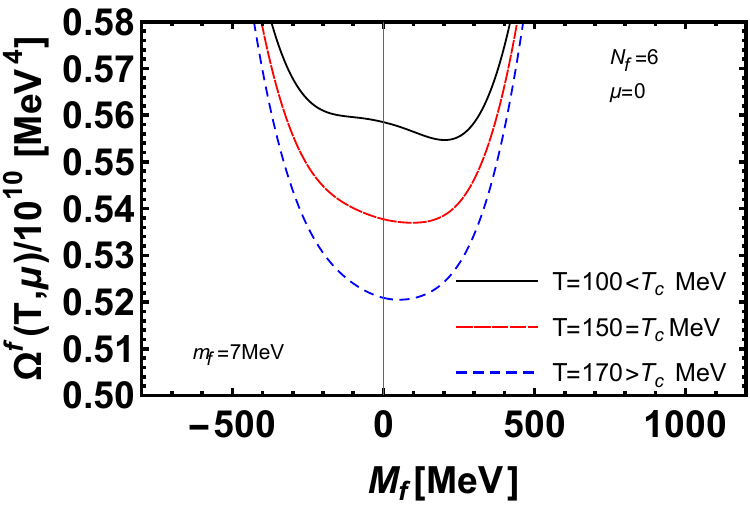}
\caption{ 
Behavior of the effective potential for at $\mu=0$ for fixed  $N_f=6$, , at $(T < T_c)$,~$(T = T_c)$, and $(T > T_c)$.  at $(T < T_c)$, the stable global minimum is located at $M_f>m_f$ and the chiral symmetry is broken. At  and above  $T=T_c$ , the stable global minimum is aligned with  $M_f=m_f$. For $(T > T_c)$, The global minimum is positioned at $M_f=m_f$, where the chiral symmetry is broken.  This plot also demonstrate  that the nature of   transition is cross-over from  chiral symmetry  broken to  restoration phase.}
\label{Fig8}
\end{center}
\end{figure}

\begin{figure}[H]
\begin{center}
\includegraphics[width=0.4\textwidth]{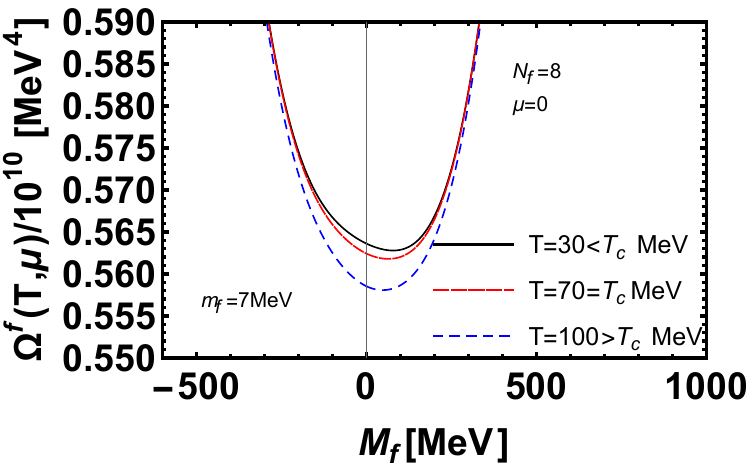}
\caption{ 
Behavior of the effective potential for at $\mu=0$ for fixed  $N_f=8$, , at $(T < T_c)$,~$(T = T_c)$, and $(T > T_c)$. For $(T < T_c)$, the chiral symmetry is broken and the global minimum is located at $M_f>m_f$. For  $T>T_c$ MeV, the  global minimum is align with $M_f= m_f$. For $(T > T_c)$ where the chiral symmetry is partially restored  and the global minimum is located at $M_f=m_f$. The nature of the transition is crossover.}
\label{Fig9}
\end{center}
\end{figure}

\begin{figure}[H]
\begin{center}
\includegraphics[width=0.4\textwidth]{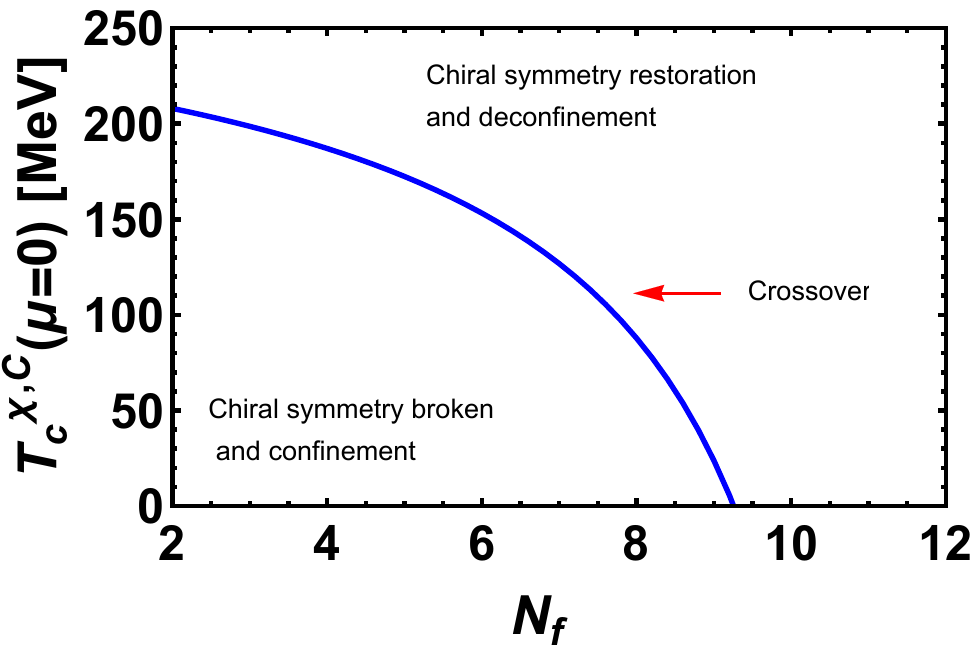}
\caption{ 
The phase diagram in the  $T^{\chi,C}_{c}$ - $N_f$  plane. This phase diagram shows that the at finite $T$ and at $\mu\rightarrow0$, the transition line between the chiral symmetry broken phase (or confinement)  and restoration (deconfinement) phase is crossover for all possible range of flavors $N_f$.}
\label{Fig10}
\end{center}
\end{figure}

\begin{figure}[H]
\begin{center}
\includegraphics[width=0.4\textwidth]{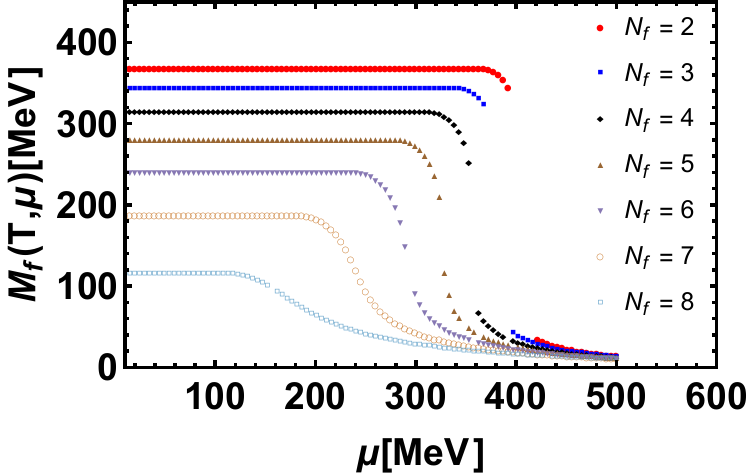}
\caption{ 
The dressed quark mass as a function of chemical potential $\mu$, for various number of flavors $N_f$, and at $T=0$. This plot shows that below $N_f=6$, the dressed mass decreases discontinuously in the chiral symmetry restoration region. While at and above, it shows the smooth decreasing behavior.}
\label{Fig11}
\end{center}
\end{figure}

\begin{figure}[H]
\begin{center}
\includegraphics[width=0.4\textwidth]{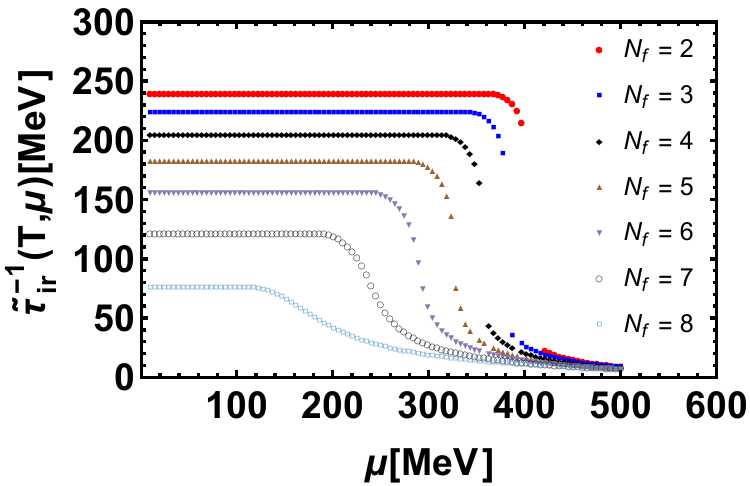}
\caption{ 
The confinement scale for various number of flavor $N_f$, and as function chemical potential $\mu$, at $T=0$. This plot indicates that below $N_f=6$, the confinement scale decreases discontinuously while at and above, it shows the smooth decreasing behavior.}
\label{Fig12}
\end{center}
\end{figure}

\begin{figure}[H]
\begin{center}
\includegraphics[width=0.48\textwidth]{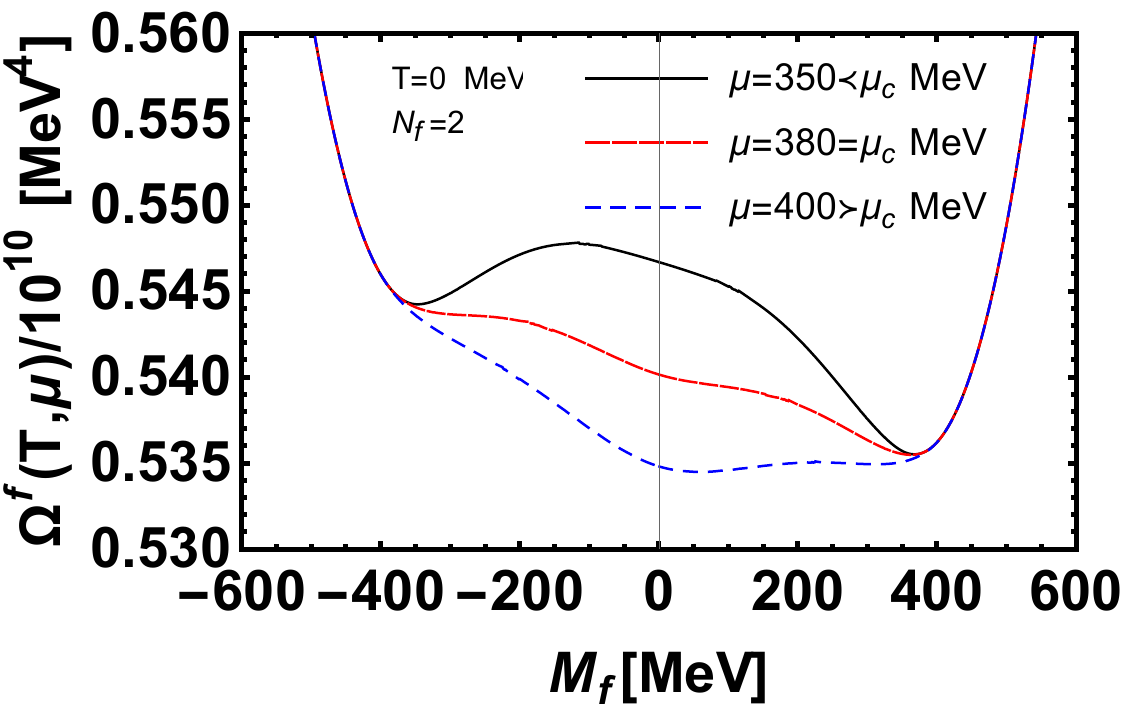}
\caption{ 
The behavior of the effective potential at $T=0$, with a fixed flavor number $ N_f=2$, is analyzed for three different chemical potentials: $\mu < \mu_c $, $ \mu = \mu_c $, and $ \mu > \mu_c $. When $\mu < \mu_c$, a stable global minimum exists at $M_f > m_f $, indicating broken chiral symmetry and a crossover transition. At the critical point $\mu = \mu_c$, the effective potential presents two global minima: one at $M_f = m_f $ and the other at $M_f > m_f$, signifying a shift from a crossover to a first-order transition. For $\mu > \mu_c$, chiral symmetry is partially restored, with the global minima shifting towards $M_f \rightarrow m_f$.}
\label{Fig13}
\end{center}
\end{figure}

\begin{figure}[H]
\begin{center}
\includegraphics[width=0.4\textwidth]{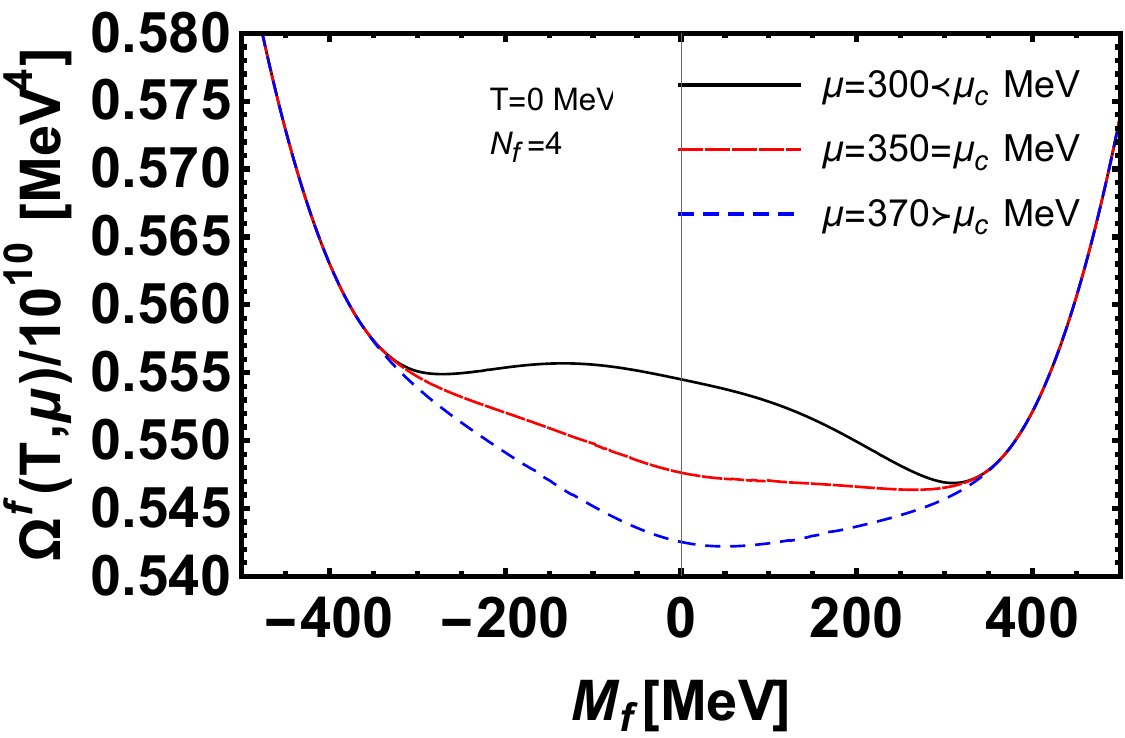}
\caption{ 
The effective potential for $N_f = 4$ reveals different behaviors at various chemical potentials: $\mu < \mu_c$, $\mu = \mu_c$, and $\mu > \mu_c$. For $\mu < \mu_c$, a stable global minimum exists at $M_f > m_f $, indicating broken chiral symmetry and a crossover transition. At the critical point $\mu = \mu_c$, the effective potential presents two global minima: one at $M_f = m_f $ and the other at $M_f > m_f$, signifying a shift from a crossover to a first-order transition. For $\mu > \mu_c$, chiral symmetry is partially restored, with the global minima shifting towards $M_f \rightarrow m_f$.}
\label{Fig14}
\end{center}
\end{figure}

\begin{figure}[H]
\begin{center}
\includegraphics[width=0.4\textwidth]{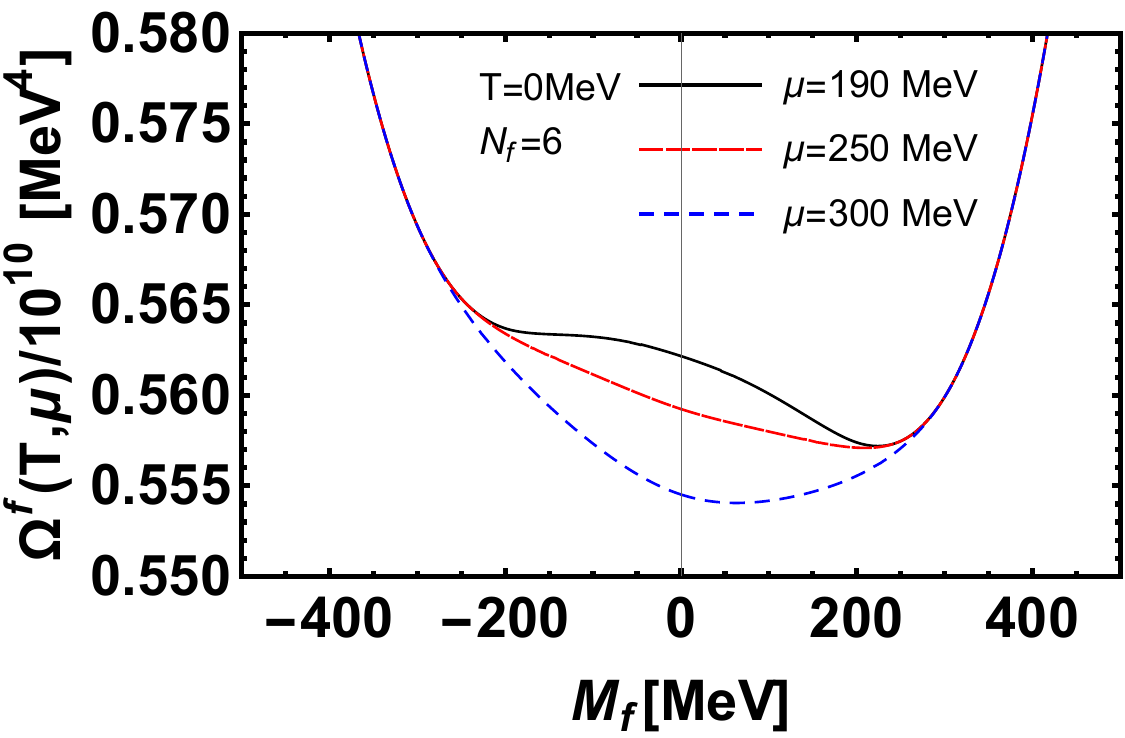}
\caption{ 
The effective potential for $N_f = 6$ reveals different behaviors at various chemical potentials: $\mu < \mu_c$, $\mu = \mu_c$, and $\mu > \mu_c$. For $\mu < \mu_c$, a stable global minimum exists at $M_f > m_f $, indicating the chiral symmetry is broken and a crossover transition. At the critical point $\mu = \mu_c$, the effective potential has now single global minim shift towards $M_f \rightarrow m_f$,  and is  crossover.  For $\mu > \mu_c$, chiral symmetry is partially restored, with the global minim positioned  at $M_f= m_f$. Thus, in this case,  the  chiral symmetry is partially restored through a cross-over  phase transition.} 
\label{Fig15}
\end{center}
\end{figure}

\begin{figure}[H]
\begin{center}
\includegraphics[width=0.4\textwidth]{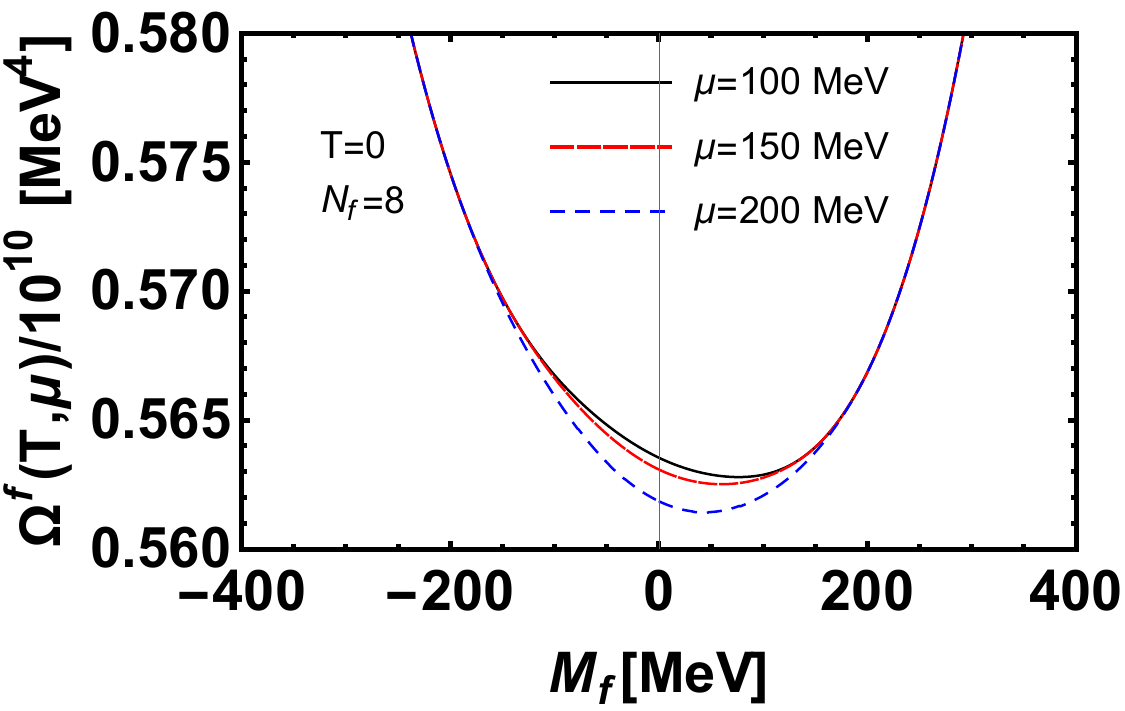}
\caption{ 
The behavior of the effective potential at $T=0$ for $N_f = 8$ is examined at three different chemical potential regimes: $\mu < \mu_c$, $\mu = \mu_c$, and $\mu > \mu_c)$. In the plot for $\mu < \mu_c$, there is a stable global minimum located at slightly higher values of $M_f > m_f$. For $\mu \geq \mu_c$, the minimum is positioned at $M_f \rightarrow m_f$.  For $\mu > \mu_c)$, the minimum is located at $M_f = m_f$ indicating that chiral symmetry is restored through a crossover phase transition. }   
\label{Fig16}
\end{center}
\end{figure}

\begin{figure}[H]
\begin{center}
\includegraphics[width=0.4\textwidth]{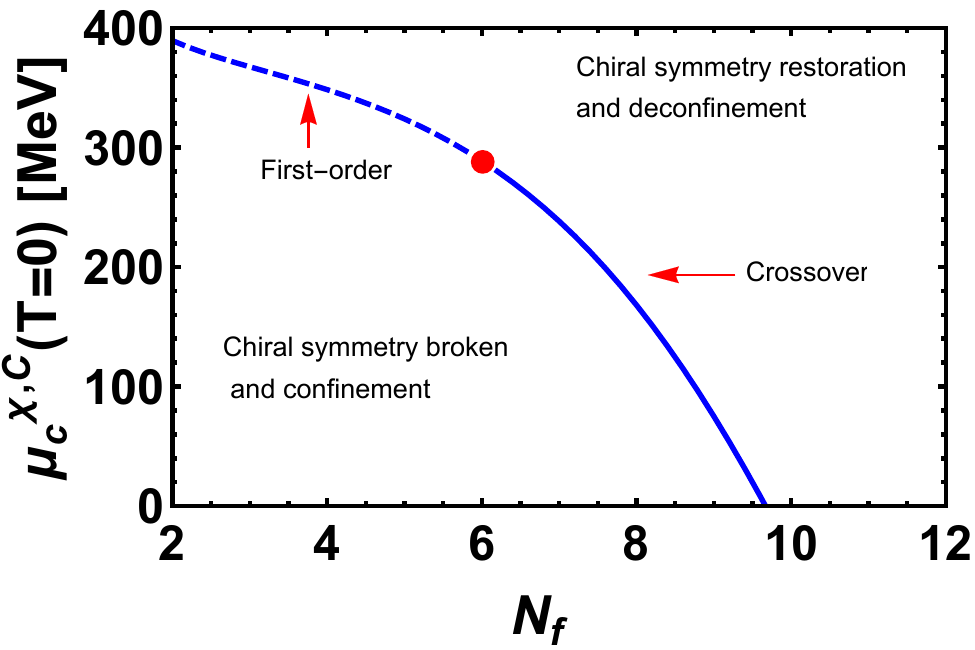}
\caption{ 
The phase diagram in the  $(\mu^{\chi,C}_{c}$ - $N_f)$  plane. This diagram shows that the at finite $\mu$ and at $T=0$, the transition line is of first-order  for $N_f<6$, while crossover for $N_f\geq6$.}
\label{Fig17}
\end{center}
\end{figure}

\begin{figure}[H]
\begin{center}
\includegraphics[width=0.4\textwidth]{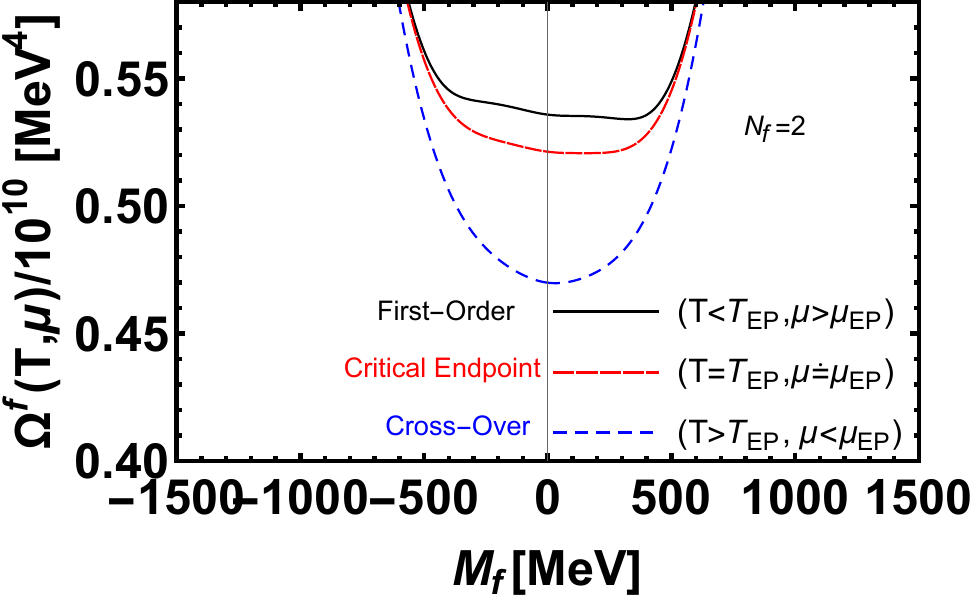}
\caption{ 
Behavior of effective potential for $N_f=2$, with bare quark mass and at finite $T$ and $\mu$, specifically  in the region for the regions $(T > T_{EP}, \mu <\mu_{EP})$, $(T = T_{EP}, \mu = \mu_{EP})$, and $(T < T_{EP}, \mu > \mu_{EP})$. This plot clearly illustrates the effective potential's behavior along the critical line, transitioning from the chiral symmetry broken phase to the chiral symmetry restoration phase.  In the crossover region $(T > T_{EP}, \mu < \mu_{EP})$, chiral symmetry is restored, resulting in a stable solution around $M_f \approx m_f$. Conversely, in the first-order transition regime $(T < T_{EP}, \mu > \mu_{EP})$, the effective potential exhibits  two  unstable global minima: one is at $M_f=m_f$ and the other is at $M_f>m_f$ . At the critical endpoint $(T = T_{EP}, \mu = \mu_{EP})$, the two global  minima  are in equilibrium  with each other.}
\label{Fig18a}
\end{center}
\end{figure}

\begin{figure}[H]
\begin{center}
\includegraphics[width=0.4\textwidth]{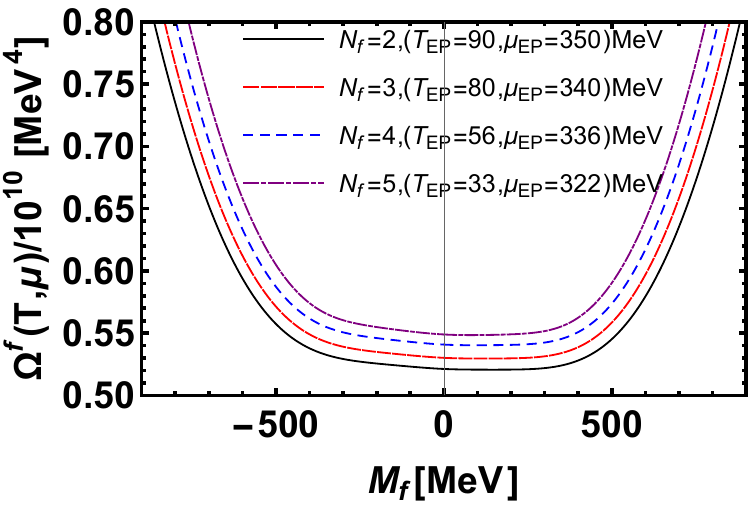}
\caption{ 
Behavior of effective potential  at the critical endpoint $(T = T_{EP}, \mu = \mu_{EP})$, for various flavors $N_f=2,3,4,5$. At the critical end point the  two global  minima  in the effective potential are in equilibrium  with each other and declined with increasing number of flavors.}
\label{Fig18b}
\end{center}
\end{figure}

\begin{figure}[H]
\begin{center}
\includegraphics[width=0.4\textwidth]{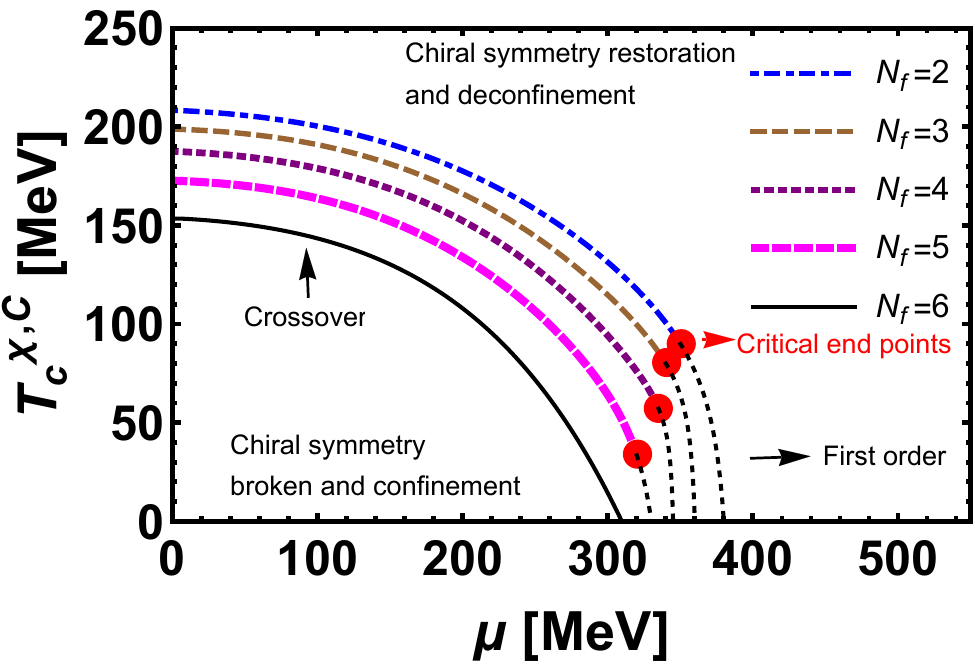}
\caption{ 
QCD phase diagram  for different $N_f=2,3,4,5,6$  values: This diagram shows the suppression of critical line among chiral symmetry broken-confinement  and chiral symmetry restoration-deconfinement transition. The coordinates of the critical end points between the crossover and first order phase transition for $N_f=2,3,4,5$ are $(T_{EP}\approx90,\mu_{EP}\approx350)$, $(T_{EP}\approx 80,\mu_{EP}\approx340)$, $(T_{EP}\approx56,\mu_{EP} \approx 336)$, $(T_{EP}\approx33,\mu_{EP}\approx322)$ MeV, respectively. However, for $N_f\geq6$  the critical line is crossover throughout the phase diagram.}
\label{Fig19}
\end{center}
\end{figure}

In this section, we focus on the numerical solution of the QCD gap equation using the $N_f$-flavor dressed contact interaction model represented by Eq.~(\ref{CI14}). We employ a specific set of model parameters: the bare light quark mass $m_f = 7$ MeV, with the infrared cutoff $\tau_{ir} = (\Lambda_{\rm QCD})^{-1} = (240\text{MeV})^{-1}$, where $\Lambda_{\rm QCD}$ denotes the typical QCD scale. Additionally, we have $\tau_{uv} =(905\text{MeV})^{-1}$, $\alpha_{ir} = 0.93\pi$, and $\mathcal{M}_g = 800$ MeV. These parameter values are sourced from~\cite{GutierrezGuerrero:2010md}, obtained by fitting to the properties of $\pi$ and $\rho$ mesons. Furthermore, electromagnetic form factors and charge radii of pseudo-scalar and scalar mesons are calculated with these parameters, as detailed in the recent review by~\cite{Hernandez-Pinto:2023yin}. Also, the first radial excitation of baryons and mass spectrum has been calculated with similar parameters~\cite{Gutierrez-Guerrero:2024him}.\\ 
In vacuum, the solution of the Eq.~(\ref{CI15a}) for  the effective potential at $T=\mu=0$, for various $N_f$ is shown in Fig.~\ref{Fig1}. This plot shows that in the chiral limit ($m_f=0$) and for $N_f=0$, the effective potential has a global minimum at $M_f \approx \pm 400$ MeV. This indicates that chiral symmetry is dynamically broken and confinement occurred, meaning that the system prefers a state where quarks acquire mass even when their bare mass is zero. In this scenario, it can be observed that the minima of the effective potential, $\Omega^{f}$, align perfectly with the solutions to the gap equation, while the trivial solution $M_f = 0$ represents a maximum. The lowest value of the positive mass reflects the global minimum of the effective potential, indicating the stable dynamical quark mass. Upon increasing the number of light quark flavors $N_f$, the global minima shifted towards the lower values of the dressed quark mass $M_f$, for example, in the case of $N_f=2$, its value is $ M_f\approx \pm 358$ MeV, and so on until $N_f\approx8$, where the global minimum shifted towards $M_f=0$, this is the stage where the chiral symmetry is restored.  In the lower panel of Fig.~\ref{Fig1}, we illustrate the behavior of the effective potential for three cases: $( N_f < N^c_f )$, $( N_f = N^c_f )$, and $( N_f > N^c_f )$. The plot reveals that for $( N_f < N^c_f )$, the global minimum occurs at nonzero values of $( M_f )$, indicating a chiral symmetry broken phase. When $( N_f = N^c_f )$, the global minimum shifts towards $( M_f \rightarrow 0 )$, signaling the beginning of chiral symmetry restoration. Lastly, for $( N_f > N^c_f )$, the global minimum is located at $( M_f = 0 )$, which confirms that the system is fully in the chiral symmetry restoration phase. \\
When the bare quark mass $m_f = 7$ MeV is included, as shown in Fig.~\ref{Fig2}, \textit{Upper panel}, the minimum shift towards $M_f \approx \pm 410$ MeV for $ N_f = 0$, and so on. This slight increase reflects the influence of the quark mass $m_f$ on the dynamics of the system. For $N_f = 2$, its value is at $M_f =\pm 368$ MeV, which is consistent with \cite{Ahmad:2023ecw}. As we increase $N_f$ to $N_f\approx 8$, the global minimum is shifted towards $M_f\rightarrow m_f$ where the chiral symmetry is partially restored (because  only the bare mass $m_f$ survives), as illustrated in Fig.~\ref{Fig2}, \textit{Upper panel}. In the \textit{Lower panel} of Fig.~\ref{Fig2}, we illustrate the behavior of the effective potential for three cases: $( N_{f} < N_{f}^{c} )$, $(N_f = N_{f}^{c} )$, and $( N_f > N_{f}^{c} )$, with $( m_f = 7 )$ MeV. The plot reveals that when $( N_f < N_{f}^{c} )$, the global minimum occurs at nonzero values of $M_f$, indicating a chiral symmetry broken phase. In the case of $N_f = N_{f}^{c} $, the global minimum shifts towards the $( M_f \rightarrow m_f )$ region, suggesting partial restoration of chiral symmetry; here, the contribution from the dressed mass due to self-energy vanishes, leaving only the bare mass contribution. For $( N_f > N_{f}^{c} )$, the global minimum is positioned at $( M_f = m_f )$, signifying that the system is in a phase of partial symmetry restoration. To understand the nature of the phase transition from chiral symmetry breaking to restoration, as well as the critical value of $N_f$, we can minimize the effective potential with respect to $M_f$ by setting $(\partial\Omega^{f}/\partial M_f = 0)$. This yields the gap equation Eq.~(\ref{CI14}) and allows us to explore the flavor gradient $(\partial_{N_f} M_f)$.\\
In the chiral limit $m_f = 0$, if the flavor-gradient diverges continuously, the transition is classified as second-order, with the effective potential exhibiting a stable solution at $(M_f = 0)$. Conversely, when a bare quark mass is present $m_f \neq 0$ and the transition remains continuous, it is characterized as a crossover, with the critical number of flavors determined at the inflection point. In this scenario, the effective potential stabilizes at $(M_f = m_f)$.
However, if the transition occurs discontinuously, it is classified as first-order, leading to unstable solutions for the effective potential in both cases, whether or not there is a bare quark mass. For a more detailed classification of phase transitions in terms of effective potential, see, for instance,~\cite{buballa2005njl}. In this first-order transition, the effective potential exhibits two minima: one at $M_f = 0) $ (in the chiral limit) or  at $M_f = m_f $ (with bare quark mass), and the other at $M_f \neq 0$ (in the chiral limit) or $M_f >m_f$ (with bare quark mass).  The criterion for a first-order phase transition is thus given by $\Omega^{f}(M_f=0, m_f) = \Omega^{f}(M_f)$, (see  for example Ref.~\cite{Kinnunen_2018} for details).\\
Overall, this analysis illustrates how the effective potential changes concerning the light-quark flavors $N_f$ and highlights the significance of dynamical symmetry breaking in QCD. The dressed quark mass $M_f$  which can be obtained by minimizing the effective potential ($\partial\Omega^{f}/ \partial M_f =0$)  and the confinement length scale, $\tilde{\tau}^{-1}_{ir}$, from which we can triggers the confinement (or deconfinement) in the chiral limit, as well as in the presence of a bare quark mass $m_f = 7$ MeV, are plotted in Fig.~\ref{Fig3}. All parameters shows a monotonically decreasing with increasing $N_f$. In the chiral limit, both the inverse of the confinement scale and the dressed mass approach zero, indicating the restoration of chiral symmetry and deconfinement. However, when a bare quark mass $m_f$ is introduced, the self-energy contribution to the dressed mass disappears, leaving only the bare quark mass contribution remains. This marks the region where chiral symmetry is restored and quarks become deconfined.\\
At finite temperature $T$  and  $\mu \rightarrow 0$, we numerically solve  Eq.~(\ref{CI20}) by minimizing it with respect to $M_f$ which gives the  gap equation Eq.~(\ref{CI20}), using a bare mass $m_f = 7$ MeV. The results, illustrated in Fig.~\ref{Fig4}, reveal how the dressed mass $M_f$ as function of  temperature $T$, varies for different values of $N_f$.  As  $T$ increases  $M_f$ monotonically  decreases and for  increasing the   value of  $N_f$, all  $M_f$ plots as a function of $T$ suppresses upon enhancing the $N_f$.  
Furthermore, the confinement length scale $\tilde{\tau}^{-1}_{ir}$, presented in Fig.~\ref{Fig5}, exhibits a similar behavior with larger $N_f$. The critical temperature $T_c$ for chiral symmetry breaking and restoration  across various values of $N_f$ can be  extracted from the inflection point of , by minimizing  the effective potential with respect to dress quark mass $(\partial \Omega^{f}/\partial M_{f}=0$), and then by taking the temperature gradient  $(\partial_{T}(\partial \Omega^{f} / \partial M_{f}=0)=\partial_{T} M_f)$ . The critical temperature for  confinement to  deconfinement  transition can be obtained from $\partial _{T}\tilde{\tau}^{-1}_{ir}$. We  thus, obtained the  critical values of temperature for various flavors, for example, for $N_f=2$ its value   is $T_c\approx207$ MeV. The nature of phase transition at finite temperature  is cross-over for all $N_f$.
In Fig.~\ref{Fig6},~\textit{Upper panel}, at finite $T$, at $\mu=0$, we present the effective potential for a fixed number of flavors, $N_f=2$, with a quark mass $m_f=7$ MeV. It is evident that as the temperature increases, the location of the global minima shifts toward lower $M_f$ values. Notably, above $T_c \approx 207$ MeV, this minimum aligns with $M_f=m_f$, indicating that above this critical temperature, chiral symmetry is partially restored while the bare quark mass persists. Conversely, below $T_c$, chiral symmetry remains broken. The  Fig.~\ref{Fig6}, ~\textit{Lower panel}, illustrates the behavior of the effective potential at three temperature regimes: $(T < T_c)$,~$(T = T_c)$, and $(T > T_c)$. It clearly shows that the effective potential has a stable minimum below $T_c$, indicating a broken chiral symmetry. At $T = T_c$, this global minimum shifts toward $M \rightarrow m_f$, where a stable solution exists, suggesting that chiral symmetry is partially restored. For $T > T_c$, the global minimum is positioned at $M_f = m_f$, where chiral symmetry is partially  restored. This plot clearly demonstrates that the effective potential provides stable solutions as it transitions from the chiral symmetry-broken phase to the restored phase across all temperature ranges, indicating that the nature of the transition is a crossover.  Similarly, we  plotted the effective potential at
$T<T_c$,~$T=T_c$ and $T>T_c$ for $N_f=4,6,8$ in the Fig.~\ref{Fig7}, Fig.~\ref{Fig8} and Fig.~\ref{Fig9}, respectively. These figures demonstrate that as the number of flavors increases, the critical temperatures decrease.
For $N_f = 8$, the situation differs from that at lower $N_f$. On one hand, the larger number of flavors restores chiral symmetry, while on the other hand, temperature plays a significant role. This results in a minimum at $M_f > m_f $  even at lower temperatures, with a critical temperature $T_c \approx 70 $ MeV where the effective potential reaches its minimum at $M_f \rightarrow m_f $. However, the transition remains a crossover.
Ultimately, we plotted the critical temperature $T^{\chi, C}_{c}$ for chiral symmetry restoration and deconfinement across various $ N_f$ values in the $T^{\chi, C}_{c} - N_f$ plane, as shown in Fig.~\ref{Fig10}. This phase diagram illustrates that, at finite temperature and as $\mu \rightarrow 0$, the transition line between the chiral symmetry broken-confinement phase and the restoration-deconfinement phase is a crossover for all flavor ranges. These findings align with studies based on the NJL model studies~\cite{Ahmad:2022hbu}.\\
At a finite chemical potential $\mu$ and at $T=0$, the dressed quark mass $M_f$ is plotted against $\mu$ for various $N_f$ in Fig.~\ref{Fig11}. This plot indicates that the dressed mass remains constant for small values of $\mu$ but experiences a sudden jump near at  $\mu=\mu_c$,with $\mu_c$  is the critical value  of quark  chemical potential.
As $N_f$ varies, $M_f$ as a function of $\mu$ declines, and at a specific number of flavors, $N_f\geq6$, the behavior of  transition changes from first-order to a smooth crossover.
A similar pattern is observed for the confinement scale $\tilde{\tau}^{-1}_{ir}$  as shown in the Fig.~\ref{Fig12}. 
Focusing on $N_f = 2$, Fig.~\ref{Fig13} presents the effective potential for three different values of quark chemical potentials.
It clearly indicates that when $\mu < \mu_c$, a stable global minimum is observed at higher values of $M_f>m_f$, signifying the broken of  chiral symmetry  and the nature of the transition is a crossover. At $\mu = \mu_c$, the transition changes from crossover to the first-order.  In the first-order transition, the effective potential exhibits an unstable solution with two global minima: one at $M_f = m_f $ and  other is at  $M_f > m_f$.  Near at the critical $\mu_c$, these two minima reaches to the state of equilibrium.  At $\mu > \mu_c$, the chiral symmetry is partially restored, with all global minima shifted to $M_f = m_f$. These behaviors are consistent with the effective potential at finite $\mu$  as discussed in~\cite{buballa2005njl}.
In contrast to a second-order (or crossover) phase transition, the location of the global minimum of the effective potential in the first-order phase transition changes discontinuously. The critical value  $\mu_c $ can be obtained from the condition $\left( \partial{\mu}\left(\partial \Omega^{f}/\partial M_{f}\right) = 0 \right) $. The critical $\mu_c$ is determined from the inflection point of $\partial{\mu}(\partial \Omega^{f}/\partial M_f=0)$. If the transition is a  first-order transition at $\mu_c$ then the derivative  changes discontinuously.
In Fig.~\ref{Fig14}, we illustrate the behavior of the effective potential for $N_f=4$. The plot reveals that their is a  single stable global minimum  positioned at  higher value of $M_f$, indicating that chiral symmetry is broken through a crossover. At $\mu = \mu_c$, there  are two  global  minima one at $M_f=m_f$ and other is located at $M_f>m_f$, and almost in equilibrium at $\mu_c$, ant at this $\mu_c$, the nature of the transition changes from crossover to the first order. For $\mu > \mu_c$, all global minima shift towards $M_f = m_f$, signifying the restoration of chiral symmetry.  For $N_f=4$, the transition from chiral symmetry breaking to restoration is a first-order.
For $N_f = 6$, as illustrated in Fig.~\ref{Fig15}, the effective potential demonstrates a smooth crossover transition from the chiral symmetry-broken phase to the restored phase, even at a finite quark chemical potential $\mu$. At $T = 0$ and for $N_f = 8$, Fig.~ \ref{Fig16} depicts the effective potential for three different values of the quark chemical potential $\mu$. These plots indicate that there is a stable minimum just above  $M_f=m_f$ for lower values of $\mu$. However, for larger values of $\mu$, the minimum coincides with $M_f = m_f$. This transition is also a crossover, and no critical endpoint for the chemical potential is identified in this scenario, as chiral symmetry is restored at $(T = \mu = 0)$ when $N_f \approx 8$.\\
We present a phase diagram depicting the critical chemical potential $\mu^{\chi,C}_{c}$ versus the number of flavors $N_f$ at temperature $T=0$, as illustrated in Fig.~\ref{Fig17}. This diagram reveals that the critical chemical potential associated with the transition from chiral symmetry breaking and confinement to chiral symmetry restoration and deconfinement decreases as the number of light quark flavors $N_f$ increases. Notably, the nature of the phase transition shifts from first-order to a crossover when $N_f\geq 6$.\\
In Fig.~\ref{Fig18a}, we demonstrate the behavior of the effective potential for $N_f=2$, considering a bare quark mass $m_f$ and at non-zero temperatures  $T$ and chemical potentials $\mu$—specifically for the regions $(T > T_{EP}, \mu <\mu_{EP})$, $(T = T_{EP}, \mu = \mu_{EP})$, and $(T < T_{EP}, \mu > \mu_{EP})$. This plot clearly illustrates the effective potential's behavior along the critical line, transitioning from the chiral symmetry broken phase to the chiral symmetry restoration phase, especially in the context of finite $T$ and $\mu$. The transition evolves from a crossover to a first-order phase transition, delineated by a critical endpoint. In the crossover region $(T > T_{EP}, \mu < \mu_{EP})$, chiral symmetry is restored, resulting in a stable solution around $M_f \approx m_f$. Conversely, in the first-order transition regime $(T < T_{EP}, \mu > \mu_{EP})$, the effective potential exhibits  with two  unstable global minima: one is at $M_f=m_f$ and the other is at $M_f>m_f$ . At the critical endpoint $(T = T_{EP}, \mu = \mu_{EP})$, the two global  minima  are in equilibrium  with each other.
Fig.~\ref{Fig18b}, illustrates the behavior of effective potential  near the critical endpoint $(T_{EP}, \mu_{EP})$ for various numbers of flavors $N_f$. This plot indicates that the critical endpoint diminishes as the number of flavors increases.\\
Finally, we draw the QCD phase diagram in the $T^{\chi,C}_c-\mu$ plane for various values of $N_f$ in Fig.~\ref{Fig19}. This phase diagram illustrates that the crossover line, which begins at the $T-$axis, does not terminate on the finite $\mu$-axis. Instead, it ends at a critical endpoint ($T_{EP},\mu_{EP}$) in the phase diagram, where its nature  changes from crossover to first-order. This line continues along the $\mu$-axis  at  $T = 0$ for lower values of $N_f$. However, as we increase $N_f$, the critical line becomes suppressed, and at $N_f = 6$ and above, it exhibits crossover behavior.
\section{ Summary and conclusions}
Our analysis  in this work based on the  Schwinger-Dyson equation, Flavor-dressed contact interaction model, in the Landau gauge, and in the rainbow ladder truncation. The expression for the gap equation is obtained using optimal  Schwinger proper time regularization and upon introducing the infrared and ultraviolet cut-offs for higher number of light quark flavors $N_f$. We have derived an expression for effective potential for large $N_f$. We further  extended this procedure at finite $T$ and $\mu$ and explored the QCD phase diagram for large $N_f$. In this work, the chiral symmetry breaking-restoration phase transition is triggered from the effective potential $\Omega^{f}$, whereas the confinement-deconfinement transition is approximated from the confinement length scale $\tilde{\tau}_{ir}$. From this study, we concluded the following. \\  
1) Chiral symmetry restoration and the deconfinement phase transition occur at and  above the critical number of light quark flavors, $N_{f}^{c}\approx 8$. This study clearly demonstrates that for lower values of $ N_f<
 N_{f}^{c}$, such as $N_f = 2$, the effective potential exhibits minima at $M_f = \pm 358$ MeV in the chiral limit. When considering a bare mass of light quarks $m_f = 7$ MeV, the minima shift to  $M = \pm 367 $ MeV, indicating a breakdown of chiral symmetry. However, at and beyond $N_f \approx 8$, the minima move closer to the bare quark mass $m_f$, because the contribution in the dressed quark mass from the self energy vanishes.\\
2) At finite $T$ and $\mu\rightarrow0$, the chiral symmetry restored and quarks becomes deconfined at some critical temperature $T^{\chi, C}_{c}\approx207$  (for $N_f=2$) where at and above,  the minima in the effective potential  positioned at  $M_f\rightarrow m_f$. Upon increasing the number of light quark flavors $N_f$, the $T^{\chi, C}_c$ reduced, the nature of phase transition is cross-over through all the ranges of $N_f$.\\
3) At finite $\mu$ and $T \rightarrow 0$, chiral symmetry is restored and quarks become deconfined through a first-order phase transition near and above $\mu^{\chi,C}_{c}\approx380$ (for $N_f=2$). The  critical temperature decreases upon  increasing the number of  flavors $N_f$,  The first-order phase transition continues up to $N_f = 5$, after which it changes from first-order to crossover.
4) At finite  $T$ and $\mu$, we have drawn the QCD phase diagram in the $(T^{\chi, C}_{c }-\mu)$ plane for various number of  $N_f$. This phase diagram indicates that their is a critical end point  $(T_{EP},\mu_{EP})$ between the crossover and first order phase transition for $(N_f=2,3,4,5)$. However, for $N_f\geq6$, no critical end point is predicted, and  the nature of the transition  is crossover.
Hence, the entire critical line between chiral symmetry breaking-confinement  and restoration-deconfinement  suppressed with the increasing number of light quark flavors.\\
The primary contribution of our work is to enhance the understanding of the QCD phase diagram with various numbers of light quark flavors at finite temperature and quark chemical potential. The chiral symmetry breaking-restoration transition is induced by the effective contact interaction potential, while the confinement-deconfinement transition is determined by the confinement scale.
Our results align well with predictions from other effective models. Therefore, one can conclude that not only do the heat bath and background fields influence the phase transition, but the number of light quarks does as well.
\section*{Acknowledgments}
I acknowledge Alfredo Raya and Adnan Bashir for their guidance, which led to the genesis of this
manuscript.  
\end{multicols}
\section{References}
\medline
\begin{multicols}{2}
%
\nocite{*}
\bibliographystyle{rmf-style}
\bibliography{Aftab}

\begin{thebibliography}{99}
\providecommand{\url}[1]{\texttt{#1}}
\providecommand{\urlprefix}{URL }
\expandafter\ifx\csname urlstyle\endcsname\relax
  \providecommand{\doi}[1]{\discretionary{}{}{}#1}\else
  \providecommand{\doi}{\discretionary{}{}{}\begingroup \urlstyle{rm}\Url}\fi

\bibitem{LSD:2014nmn}
T.~Appelquist et~al.,
\newblock {Lattice simulations with eight flavors of domain wall fermions in SU(3) gauge theory},
\newblock Phys. Rev. D 90 (2014) 114502,
\newblock \url{10.1103/PhysRevD.90.114502}

\bibitem{Hayakawa:2010yn}
M.~Hayakawa, et~al.,
\newblock {Running coupling constant of ten-flavor QCD with the Schr\"odinger functional method},
\newblock Phys. Rev. D 83 (2011) 074509,
\newblock \url{10.1103/PhysRevD.83.074509}

\bibitem{Cheng:2013eu}
A.~Cheng, et~al.,
\newblock {Scale-dependent mass anomalous dimension from Dirac eigenmodes},
\newblock JHEP 07 (2013) 061,
\newblock \url{10.1007/JHEP07(2013)061}

\bibitem{Hasenfratz:2016dou}
A.~Hasenfratz and D.~Schaich,
\newblock {Nonperturbative $\beta$ function of twelve-flavor SU(3) gauge theory},
\newblock JHEP 02 (2018) 132,
\newblock \url{10.1007/JHEP02(2018)132}

\bibitem{LatticeStrongDynamics:2018hun}
T.~Appelquist et~al.,
\newblock {Nonperturbative investigations of SU(3) gauge theory with eight dynamical flavors},
\newblock Phys. Rev. D 99 (2019) 014509,
\newblock \url{10.1103/PhysRevD.99.014509}

\bibitem{bashir2013qcd}
A.~Bashir, A.~Raya, and J.~Rodriguez-Quintero,
\newblock QCD: restoration of chiral symmetry and deconfinement for large N f,
\newblock Physical Review D 88 (2013) 054003

\bibitem{Ahmad:2020jzn}
A.~Ahmad, et~al.,
\newblock {Flavor, temperature and magnetic field dependence of the QCD phase diagram: magnetic catalysis and its inverse},
\newblock J. Phys. G 48 (2021) 075002,
\newblock \url{10.1088/1361-6471/abd88f}

\bibitem{Ahmad:2020ifp}
A.~Ahmad,
\newblock {Chiral symmetry restoration and deconfinement in the contact interaction model of quarks with parallel electric and magnetic fields},
\newblock Chin. Phys. C 45 (2021) 073109,
\newblock \url{10.1088/1674-1137/abfb5f}

\bibitem{Ahmad:2022hbu}
A.~Ahmad and A.~Murad,
\newblock {Color-flavor dependence of the Nambu-Jona-Lasinio model and QCD phase diagram},
\newblock Chin. Phys. C 46 (2022) 083109,
\newblock \url{10.1088/1674-1137/ac6cd8}

\bibitem{Ahmad:2023mqg}
A.~Ahmad and A.~Farooq,
\newblock {Schwinger Pair Production in QCD from Flavor-Dependent Contact Interaction Model of Quarks},
\newblock Braz. J. Phys. 54 (2024) 212,
\newblock \url{10.1007/s13538-024-01581-0}

\bibitem{Appelquist:1998rb}
T.~Appelquist, et~al.,
\newblock {The Phase structure of an SU(N) gauge theory with N(f) flavors},
\newblock Phys. Rev. D 58 (1998) 105017,
\newblock \url{10.1103/PhysRevD.58.105017}

\bibitem{Hopfer:2014zna}
M.~Hopfer, C.~S. Fischer, and R.~Alkofer,
\newblock {Running coupling in the conformal window of large-Nf QCD},
\newblock JHEP 11 (2014) 035,
\newblock \url{10.1007/JHEP11(2014)035}

\bibitem{Doff:2016jzk}
A.~Doff and A.~A. Natale,
\newblock {Anomalous mass dimension in multiflavor QCD},
\newblock Phys. Rev. D 94 (2016) 076005,
\newblock \url{10.1103/PhysRevD.94.076005}

\bibitem{Binosi:2016xxu}
D.~Binosi, C.~D. Roberts, and J.~Rodriguez-Quintero,
\newblock {Scale-setting, flavor dependence, and chiral symmetry restoration},
\newblock Phys. Rev. D 95 (2017) 114009,
\newblock \url{10.1103/PhysRevD.95.114009}

\bibitem{Evans:2020ztq}
N.~Evans and K.~S. Rigatos,
\newblock {Chiral symmetry breaking and confinement: separating the scales},
\newblock Phys. Rev. D 103 (2021) 094022,
\newblock \url{10.1103/PhysRevD.103.094022}

\bibitem{Zierler:2023qvz}
F.~Zierler and R.~Alkofer,
\newblock {Dependence of the Landau gauge ghost-gluon-vertex on the number of flavors},
\newblock Phys. Rev. D 109 (2024) 074024,
\newblock \url{10.1103/PhysRevD.109.074024}

\bibitem{Politzer:1973fx}
H.~D. Politzer,
\newblock {Reliable Perturbative Results for Strong Interactions?},
\newblock Phys. Rev. Lett. 30 (1973) 1346,
\newblock \url{10.1103/PhysRevLett.30.1346}

\bibitem{Caswell:1974gg}
W.~E. Caswell,
\newblock {Asymptotic Behavior of Nonabelian Gauge Theories to Two Loop Order},
\newblock Phys. Rev. Lett. 33 (1974) 244,
\newblock \url{10.1103/PhysRevLett.33.244}

\bibitem{Banks:1981nn}
T.~Banks and A.~Zaks,
\newblock {On the Phase Structure of Vector-Like Gauge Theories with Massless Fermions},
\newblock Nucl. Phys. B 196 (1982) 189,
\newblock \url{10.1016/0550-3213(82)90035-9}

\bibitem{gies2006chiral}
H.~Gies and J.~Jaeckel,
\newblock Chiral phase structure of QCD with many flavors,
\newblock The European Physical Journal C-Particles and Fields 46 (2006) 433

\bibitem{Appelquist:2007hu}
T.~Appelquist, G.~T. Fleming, and E.~T. Neil,
\newblock {Lattice study of the conformal window in QCD-like theories},
\newblock Phys. Rev. Lett. 100 (2008) 171607,
\newblock \url{10.1103/PhysRevLett.100.171607}

\bibitem{hasenfratz2010conformal}
A.~Hasenfratz,
\newblock Conformal or walking? Monte Carlo renormalization group studies of S U (3) gauge models with fundamental fermions,
\newblock Physical Review D 82 (2010) 014506

\bibitem{Aoki:2011rrd}
Y.~Aoki, et~al.,
\newblock {Many flavor QCD as exploration of the walking behavior with the approximate IR fixed point},
\newblock PoS LATTICE2011 (2011) 080,
\newblock \url{10.22323/1.139.0080}

\bibitem{Appelquist:2009ty}
T.~Appelquist, G.~T. Fleming, and E.~T. Neil,
\newblock {Lattice Study of Conformal Behavior in SU(3) Yang-Mills Theories},
\newblock Phys. Rev. D 79 (2009) 076010,
\newblock \url{10.1103/PhysRevD.79.076010}

\bibitem{LSD:2009yru}
T.~Appelquist et~al.,
\newblock {Toward TeV Conformality},
\newblock Phys. Rev. Lett. 104 (2010) 071601,
\newblock \url{10.1103/PhysRevLett.104.071601}

\bibitem{Zhitnitsky:2013wfa}
A.~R. Zhitnitsky,
\newblock {Conformal window in QCD for large numbers of colors and flavors},
\newblock Nucl. Phys. A 921 (2014) 1,
\newblock \url{10.1016/j.nuclphysa.2013.10.011}

\bibitem{Lee:2020ihn}
J.-W. Lee,
\newblock {Conformal window from conformal expansion},
\newblock Phys. Rev. D 103 (2021) 076006,
\newblock \url{10.1103/PhysRevD.103.076006}

\bibitem{LatticeStrongDynamics:2023bqp}
R.~C. Brower et~al.,
\newblock {Light scalar meson and decay constant in SU(3) gauge theory with eight dynamical flavors},
\newblock Phys. Rev. D 110 (2024) 054501,
\newblock \url{10.1103/PhysRevD.110.054501}

\bibitem{Ahmad:2024emu}
A.~Ahmad and M.~Khan,
\newblock {$\pi$- and $K$-meson properties for large $N_f$ and $N_c$},
\newblock arXiv preprint arXiv:2401.11186  (2024)

\bibitem{rischke1988phase}
D.~Rischke, et~al.,
\newblock Phase transition from hadron gas to quark-gluon plasma: influence of the stiffness of the nuclear equation of state,
\newblock Journal of Physics G: Nuclear Physics 14 (1988) 191

\bibitem{mclerran2009quarkyonic}
L.~McLerran,
\newblock Quarkyonic matter and the revised phase diagram of QCD,
\newblock Nuclear Physics A 830 (2009) 709c

\bibitem{McLerran:2018hbz}
L.~McLerran and S.~Reddy,
\newblock {Quarkyonic Matter and Neutron Stars},
\newblock Phys. Rev. Lett. 122 (2019) 122701,
\newblock \url{10.1103/PhysRevLett.122.122701}

\bibitem{Bluhm:2024uhj}
M.~Bluhm, et~al.,
\newblock {Quark saturation in the QCD phase diagram}  (2024)

\bibitem{shao2011evolution}
G.-y. Shao,
\newblock Evolution of proto-neutron stars with the hadron--quark phase transition,
\newblock Physics Letters B 704 (2011) 343

\bibitem{Adhikari:2024bfa}
P.~Adhikari et~al.,
\newblock {Strongly interacting matter in extreme magnetic fields}  (2024)

\bibitem{Barrois:1977xd}
B.~C. Barrois,
\newblock {Superconducting Quark Matter},
\newblock Nucl. Phys. B 129 (1977) 390,
\newblock \url{10.1016/0550-3213(77)90123-7}

\bibitem{Casalbuoni:1999zi}
R.~Casalbuoni and R.~Gatto,
\newblock {The color flavor locking phase at T not equal to 0: Exact results at order T**2},
\newblock Phys. Lett. B 469 (1999) 213,
\newblock \url{10.1016/S0370-2693(99)01274-5}

\bibitem{Rajagopal:1999cp}
K.~Rajagopal,
\newblock {Mapping the QCD phase diagram},
\newblock Nucl. Phys. A 661 (1999) 150,
\newblock \url{10.1016/S0375-9474(99)85017-9}

\bibitem{Arslandok:2023utm}
M.~Arslandok et~al.,
\newblock {Hot QCD White Paper}  (2023)

\bibitem{durante2019all}
M.~Durante, et~al.,
\newblock All the fun of the FAIR: fundamental physics at the facility for antiproton and ion research,
\newblock Physica Scripta 94 (2019) 033001

\bibitem{Kolesnikov:2020qfw}
V.~I. Kolesnikov, et~al.,
\newblock {Progress in the construction of the NICA accelerator complex},
\newblock Phys. Scripta 95 (2020) 094001,
\newblock \url{10.1088/1402-4896/aba665}

\bibitem{Aoki:2006we}
Y.~Aoki, et~al.,
\newblock {The Order of the quantum chromodynamics transition predicted by the standard model of particle physics},
\newblock Nature 443 (2006) 675,
\newblock \url{10.1038/nature05120}

\bibitem{Cheng:2006qk}
M.~Cheng et~al.,
\newblock {The Transition temperature in QCD},
\newblock Phys. Rev. D 74 (2006) 054507,
\newblock \url{10.1103/PhysRevD.74.054507}

\bibitem{Bhattacharya:2014ara}
T.~Bhattacharya et~al.,
\newblock {QCD Phase Transition with Chiral Quarks and Physical Quark Masses},
\newblock Phys. Rev. Lett. 113 (2014) 082001,
\newblock \url{10.1103/PhysRevLett.113.082001}

\bibitem{deForcrand:2014tha}
P.~de~Forcrand, et~al.,
\newblock {Lattice QCD Phase Diagram In and Away from the Strong Coupling Limit},
\newblock Phys. Rev. Lett. 113 (2014) 152002,
\newblock \url{10.1103/PhysRevLett.113.152002}

\bibitem{HotQCD:2018pds}
A.~Bazavov et~al.,
\newblock {Chiral crossover in QCD at zero and non-zero chemical potentials},
\newblock Phys. Lett. B 795 (2019) 15,
\newblock \url{10.1016/j.physletb.2019.05.013}

\bibitem{Borsanyi:2020fev}
S.~Borsanyi, et~al.,
\newblock {QCD Crossover at Finite Chemical Potential from Lattice Simulations},
\newblock Phys. Rev. Lett. 125 (2020) 052001,
\newblock \url{10.1103/PhysRevLett.125.052001}

\bibitem{Guenther:2020jwe}
J.~N. Guenther,
\newblock {Overview of the QCD phase diagram: Recent progress from the lattice},
\newblock Eur. Phys. J. A 57 (2021) 136,
\newblock \url{10.1140/epja/s10050-021-00354-6}

\bibitem{Borsanyi:2025lim}
S.~Bors\'anyi, et~al.,
\newblock {Chiral versus deconfinement properties of the QCD crossover: Differences in the volume and chemical potential dependence from the lattice},
\newblock Phys. Rev. D 111 (2025) 014506,
\newblock \url{10.1103/PhysRevD.111.014506}

\bibitem{Qin:2010nq}
S.-x. Qin, et~al.,
\newblock {Phase diagram and critical endpoint for strongly-interacting quarks},
\newblock Phys. Rev. Lett. 106 (2011) 172301,
\newblock \url{10.1103/PhysRevLett.106.172301}

\bibitem{Fischer:2011mz}
C.~S. Fischer, J.~Luecker, and J.~A. Mueller,
\newblock {Chiral and deconfinement phase transitions of two-flavour QCD at finite temperature and chemical potential},
\newblock Phys. Lett. B 702 (2011) 438,
\newblock \url{10.1016/j.physletb.2011.07.039}

\bibitem{gutierrez2014qcd}
E.~Guti{\'e}rrez, et~al.,
\newblock The QCD phase diagram from Schwinger--Dyson equations,
\newblock Journal of Physics G: Nuclear and Particle Physics 41 (2014) 075002

\bibitem{Eichmann:2015kfa}
G.~Eichmann, C.~S. Fischer, and C.~A. Welzbacher,
\newblock {Baryon effects on the location of QCD\textquoteright{}s critical end point},
\newblock Phys. Rev. D 93 (2016) 034013,
\newblock \url{10.1103/PhysRevD.93.034013}

\bibitem{Ahmad:2015cgh}
A.~Ahmad, et~al.,
\newblock {QCD Phase Diagram and the Constant Mass Approximation},
\newblock J. Phys. Conf. Ser. 651 (2015) 012018,
\newblock \url{10.1088/1742-6596/651/1/012018}

\bibitem{Gao:2016qkh}
F.~Gao and Y.-x. Liu,
\newblock {QCD phase transitions via a refined truncation of Dyson-Schwinger equations},
\newblock Phys. Rev. D 94 (2016) 076009,
\newblock \url{10.1103/PhysRevD.94.076009}

\bibitem{Ahmad:2016iez}
A.~Ahmad and A.~Raya,
\newblock Inverse magnetic catalysis and confinement within a contact interaction model for quarks,
\newblock Journal of Physics G: Nuclear and Particle Physics 43 (2016) 065002

\bibitem{Fischer:2018sdj}
C.~S. Fischer,
\newblock {QCD at finite temperature and chemical potential from Dyson\textendash{}Schwinger equations},
\newblock Prog. Part. Nucl. Phys. 105 (2019) 1,
\newblock \url{10.1016/j.ppnp.2019.01.002}

\bibitem{Shi:2020uyb}
C.~Shi, et~al.,
\newblock {Chiral transition and the chiral charge density of the hot and dense QCD matter},
\newblock JHEP 06 (2020) 122,
\newblock \url{10.1007/JHEP06(2020)122}

\bibitem{klevansky1992nambu}
S.~Klevansky,
\newblock The Nambu—Jona-Lasinio model of quantum chromodynamics,
\newblock Reviews of Modern Physics 64 (1992) 649

\bibitem{buballa2005njl}
M.~Buballa,
\newblock NJL-model analysis of dense quark matter,
\newblock Physics Reports 407 (2005) 205

\bibitem{costa2010phase}
P.~Costa, et~al.,
\newblock Phase diagram and critical properties within an effective model of QCD: the Nambu--Jona-Lasinio model coupled to the Polyakov loop,
\newblock Symmetry 2 (2010) 1338

\bibitem{Ayala:2011vs}
A.~Ayala, et~al.,
\newblock {QCD phase diagram from finite energy sum rules},
\newblock Phys. Rev. D 84 (2011) 056004,
\newblock \url{10.1103/PhysRevD.84.056004}

\bibitem{Marquez:2015bca}
F.~Marquez, et~al.,
\newblock {The dual quark condensate in local and nonlocal NJL models: an order parameter for deconfinement?},
\newblock Phys. Lett. B747 (2015) 529,
\newblock \url{10.1016/j.physletb.2015.06.031}

\bibitem{Ayala:2017gek}
A.~Ayala, et~al.,
\newblock {Locating the critical end point using the linear sigma model coupled to quarks},
\newblock EPJ Web Conf. 172 (2018) 02003,
\newblock \url{10.1051/epjconf/201817202003}

\bibitem{Ayala:2021nhx}
A.~Ayala, et~al.,
\newblock {QCD phase diagram in a magnetized medium from the chiral symmetry perspective: the linear sigma model with quarks and the Nambu\textendash{}Jona-Lasinio model effective descriptions},
\newblock Eur. Phys. J. A 57 (2021) 234,
\newblock \url{10.1140/epja/s10050-021-00534-4}

\bibitem{Ahmad:2023ecw}
A.~Ahmad, M.~Azher, and A.~Raya,
\newblock {Robust features of a QCD phase diagram through a contact interaction model for quarks: a view from the effective potential},
\newblock Eur. Phys. J. A 59 (2023) 252,
\newblock \url{10.1140/epja/s10050-023-01169-3}

\bibitem{Sasaki:2007qh}
C.~Sasaki, B.~Friman, and K.~Redlich,
\newblock {Chiral phase transition in the presence of spinodal decomposition},
\newblock Phys. Rev. D 77 (2008) 034024,
\newblock \url{10.1103/PhysRevD.77.034024}

\bibitem{Costa:2008yh}
P.~Costa, M.~C. Ruivo, and C.~A. de~Sousa,
\newblock {Thermodynamics and critical behavior in the Nambu-Jona-Lasinio model of QCD},
\newblock Phys. Rev. D 77 (2008) 096001,
\newblock \url{10.1103/PhysRevD.77.096001}

\bibitem{Fu:2007xc}
W.-j. Fu, Z.~Zhang, and Y.-x. Liu,
\newblock {2+1 flavor Polyakov-Nambu-Jona-Lasinio model at finite temperature and nonzero chemical potential},
\newblock Phys. Rev. D 77 (2008) 014006,
\newblock \url{10.1103/PhysRevD.77.014006}

\bibitem{Abuki:2008nm}
H.~Abuki, et~al.,
\newblock {Chiral crossover, deconfinement and quarkyonic matter within a Nambu-Jona Lasinio model with the Polyakov loop},
\newblock Phys. Rev. D 78 (2008) 034034,
\newblock \url{10.1103/PhysRevD.78.034034}

\bibitem{Loewe:2013zaa}
M.~Loewe, F.~Marquez, and C.~Villavicencio,
\newblock {The nNJL model with a fractional Lorentzian regulator in the real time formalism},
\newblock Phys. Rev. D 88 (2013) 056004,
\newblock \url{10.1103/PhysRevD.88.056004}

\bibitem{Kovacs:2007sy}
P.~Kovacs and Z.~Szep,
\newblock {Influence of the isospin and hypercharge chemical potentials on the location of the CEP in the mu(B) - T phase diagram of the SU(3)(L) x SU(3)(R) chiral quark model},
\newblock Phys. Rev. D 77 (2008) 065016,
\newblock \url{10.1103/PhysRevD.77.065016}

\bibitem{Schaefer:2007pw}
B.-J. Schaefer, J.~M. Pawlowski, and J.~Wambach,
\newblock {The Phase Structure of the Polyakov--Quark-Meson Model},
\newblock Phys. Rev. D 76 (2007) 074023,
\newblock \url{10.1103/PhysRevD.76.074023}

\bibitem{Bazavov:2011nk}
A.~Bazavov et~al.,
\newblock {The chiral and deconfinement aspects of the QCD transition},
\newblock Phys. Rev. D 85 (2012) 054503,
\newblock \url{10.1103/PhysRevD.85.054503}

\bibitem{fodor2002lattice}
Z.~Fodor and S.~D. Katz,
\newblock Lattice determination of the critical point of QCD at finite T and $\mu$,
\newblock Journal of High Energy Physics 2002 (2002) 014

\bibitem{gavai2005critical}
R.~V. Gavai and S.~Gupta,
\newblock On the critical end point of QCD,
\newblock Physical Review D 71 (2005) 114014

\bibitem{li2009study}
A.~Li, et~al.,
\newblock Study of QCD critical point using canonical ensemble method,
\newblock Nuclear Physics A 830 (2009) 633c

\bibitem{deForcrand:2006ec}
P.~de~Forcrand and S.~Kratochvila,
\newblock {Finite density QCD with a canonical approach},
\newblock Nucl. Phys. B Proc. Suppl. 153 (2006) 62,
\newblock \url{10.1016/j.nuclphysbps.2006.01.007}

\bibitem{Wang:2013wk}
K.-l. Wang, et~al.,
\newblock {Baryon and meson screening masses},
\newblock Phys. Rev. D87 (2013) 074038,
\newblock \url{10.1103/PhysRevD.87.074038}

\bibitem{Boucaud:2011ug}
P.~Boucaud, et~al.,
\newblock {The Infrared Behaviour of the Pure Yang-Mills Green Functions},
\newblock Few Body Syst. 53 (2012) 387,
\newblock \url{10.1007/s00601-011-0301-2}

\bibitem{Solis:2019fzm}
E.~L. Solis, et~al.,
\newblock {Quark propagator in Minkowski space},
\newblock Few Body Syst. 60 (2019) 49,
\newblock \url{10.1007/s00601-019-1517-9}

\bibitem{Ebert:1996vx}
D.~Ebert, T.~Feldmann, and H.~Reinhardt,
\newblock {Extended NJL model for light and heavy mesons without q - anti-q thresholds},
\newblock Phys. Lett. B388 (1996) 154,
\newblock \url{10.1016/0370-2693(96)01158-6}

\bibitem{GutierrezGuerrero:2010md}
L.~X. Gutierrez-Guerrero, et~al.,
\newblock {Pion form factor from a contact interaction},
\newblock Phys. Rev. C81 (2010) 065202,
\newblock \url{10.1103/PhysRevC.81.065202}

\bibitem{Roberts:2011cf}
H.~L.~L. Roberts, et~al.,
\newblock {Masses of ground and excited-state hadrons},
\newblock Few Body Syst. 51 (2011) 1,
\newblock \url{10.1007/s00601-011-0225-x}

\bibitem{Roberts:2011wy}
H.~L.~L. Roberts, et~al.,
\newblock {pi- and rho-mesons, and their diquark partners, from a contact interaction},
\newblock Phys. Rev. C83 (2011) 065206,
\newblock \url{10.1103/PhysRevC.83.065206}

\bibitem{Hernandez-Pinto:2023yin}
R.~J. Hern\'andez-Pinto, et~al.,
\newblock {Electromagnetic form factors and charge radii of pseudoscalar and scalar mesons: A comprehensive contact interaction analysis},
\newblock Phys. Rev. D 107 (2023) 054002,
\newblock \url{10.1103/PhysRevD.107.054002}

\bibitem{Gutierrez-Guerrero:2024him}
L.~X. Guti\'errez-Guerrero, et~al.,
\newblock {First radial excitations of baryons in a contact interaction: Mass spectrum},
\newblock Phys. Rev. D 110 (2024) 074015,
\newblock \url{10.1103/PhysRevD.110.074015}

\bibitem{Kinnunen_2018}
J.~J. Kinnunen, et~al.,
\newblock The Fulde–Ferrell–Larkin–Ovchinnikov state for ultracold fermions in lattice and harmonic potentials: a review,
\newblock Reports on Progress in Physics 81 (2018) 046401,
\newblock \url{10.1088/1361-6633/aaa4ad}

\end{thebibliography}
%
%
%

%
%
\end{multicols}
\end{document}